\documentclass[aps,prd,preprint,superscriptaddress,amsmath,amssymb,showpacs]{revtex4-1}
\usepackage{todonotes}
\usepackage{dcolumn}
\usepackage{graphicx}
\usepackage{float}
\usepackage{physics}
\usepackage[colorlinks=true,allcolors=blue]{hyperref}
\usepackage{epstopdf}
\usepackage{changes}
\usepackage{ulem}
\usepackage{xcolor}
\usepackage{placeins}
\usepackage{hyperref}
\normalsize  
\usepackage{booktabs}  
\usepackage{caption}
\usepackage{capt-of}
\usepackage{subcaption}
\begin{document}
    \title{Heavy Quarkonium Spectrum and Decay Constants from a Neural-Network-Based Holographic Model}
    
    \author{Yu Zhang}
    \affiliation{School of Nuclear Science and Technology, University of South China Hengyang, No 28, West Changsheng Road, Hengyang City, Hunan Province, China.}
    \affiliation{Key Laboratory of Advanced Nuclear Energy Design and Safety, Ministry of Education, Hengyang, 421001, China}
    
    \author{Xun Chen}
    \email{chenxun@usc.edu.cn}
    \affiliation{School of Nuclear Science and Technology, University of South China Hengyang, No 28, West Changsheng Road, Hengyang City, Hunan Province, China.}
    \affiliation{Key Laboratory of Advanced Nuclear Energy Design and Safety, Ministry of Education, Hengyang, 421001, China}
    \affiliation{INFN --- Istituto Nazionale di Fisica Nucleare --- Sezione di Bari Via Orabona 4, 70125, Bari, Italy}
    
    \author{Miguel Angel Martin Contreras}
    \email{miguelangel.martin@usc.edu.cn}
    \affiliation{School of Nuclear Science and Technology, University of South China Hengyang, No 28, West Changsheng Road, Hengyang City, Hunan Province, China.}
    \affiliation{Key Laboratory of Advanced Nuclear Energy Design and Safety, Ministry of Education, Hengyang, 421001, China}

    \date{\today}
    
    \begin{abstract}  
        We present a data-driven inverse construction of the dilaton field in a bottom-up AdS/QCD description of heavy vector quarkonia. Instead of adopting an \emph{ad hoc} analytic ansatz, we use a multilayer perceptron to learn \(\Phi'(z)\) as a smooth function of the holographic coordinate, with \(\Phi(0)=0\) imposed to ensure ultraviolet consistency. The dilaton and its derivatives obtained by automatic differentiation generate the holographic potential \(U(z)\), and the associated Schrödinger-like equation is discretized and diagonalized to extract the low-lying eigenmodes. Masses and decay constants are then evaluated from the eigenvalues and the near-boundary behavior of the bulk-to-boundary modes. Training on PDG data for charmonium and bottomonium yields a non-quadratic dilaton profile that resolves the longstanding difficulty of simultaneously reproducing both the heavy-quarkonium spectrum and the monotonic suppression of leptonic decay constants with radial excitation. The combined fit achieves RMS deviations of \(1.26\%\) (charmonium) and \(3.32\%\) (bottomonium). This work establishes neural-network reconstruction as a flexible tool for holographic modeling and provides a basis for future extensions incorporating additional channels, lattice constraints, or finite-temperature backgrounds.
    \end{abstract}

    \maketitle
    
\section{Introduction} 
Quantum Chromodynamics (QCD) is the fundamental theory that describes the strong interaction. However, in the low-energy and strongly coupled regime, analytical solutions to QCD become extremely difficult to obtain, and traditional perturbative methods are no longer effective. This limitation has motivated the development of various non-perturbative approaches, among which the AdS/QCD model \cite{Polchinski:2001tt, BoschiFilho:2002vd, Erlich:2005qh, Brodsky:2007hb} based on the holographic principle has attracted considerable attention in recent years. The idea of AdS/QCD originates from the AdS/CFT correspondence, which reveals a profound connection between strongly coupled gauge theories and weakly coupled gravity theories \cite{Maldacena:1997re, Aharony:1999ti, Gubser:1998bc}. Depending on the construction scheme, AdS/QCD models can be classified into top-down and bottom-up approaches: the former are derived from string theory, where probe branes are introduced into the bulk geometry to set the energy scale and break conformal symmetry \cite{Karch:2002sh, Sakai:2004cn, Sakai:2005yt}; the latter directly implement deformations in the five-dimensional spacetime to reproduce hadronic properties observed in experiments. Within the bottom-up framework, the hard-wall \cite{Polchinski:2001tt, BoschiFilho:2002vd, BoschiFilho:2002ta} and soft-wall \cite{Karch:2006pv} models are the most widely used. In particular, the soft-wall model is well-suited for studying meson radial excitations, as it naturally reproduces the linear Regge trajectories of hadron masses. These models have been extensively applied to the study of meson spectroscopy \cite{Karch:2006pv, Grigoryan:2007my, Erdmenger:2007cm, Colangelo:2008us, BallonBayona:2009ar, Cotrone:2010fy, Li:2013oda,Wen:2024hgu,Cao:2021tcr,ThomasArun:2025uyi} and heavy quarkonia \cite{Kim:2007zzm, Grigoryan:2010pj, Li:2015zda, Braga:2015jca}. The latter have introduced quadratic-like dilaton fields to fit heavy vector quarkonia with linear radial Regge trajectories. In addition, Regge trajectory methods have recently been extended to more exotic hadronic systems, such as doubly heavy triquarks $((Qq)\bar{Q}')$, providing new spectroscopic predictions for these states~\cite{Liu:2026triquark}.

However, for heavy quarkonium systems, the linear radial Regge trajectory hypothesis breaks down, according to Bethe-Salpeter Theory \cite{Chen:2018bbr, Chen:2021kfw, Chen:2023djq}, implying Regge trajectories can be parametrized as $M_n^2=a(n+b)^\nu$. This non-linearity, at the holographic level, can be translated into the dilaton field $\Phi(z)$ if a non-quadratic version is considered, as it was introduced in \cite{MartinContreras:2020cyg}. Recently, this dilaton phenomenological \emph{ansatz} was derived from WKB analysis \cite{MartinContreras:2025wnh}.

On the purely phenomenological side, heavy quarkonia, including charmonium ($c\,\bar{c}$) and bottomonium ($b\,\bar{b}$), play a central role in hadron physics due to their large quark masses, which simplify their internal dynamics. These systems provide an effective framework for investigating the non-perturbative regime of QCD. Heavy quarkonia connect perturbative calculations with strong-coupling phenomena, enabling rigorous tests of fundamental QCD concepts such as confinement and mass generation. Systematic analyses of their mass spectra and decay constants yield valuable information about the quark-antiquark potential. Furthermore, the well-characterized properties of heavy quarkonia serve as essential benchmarks for theoretical models, including lattice QCD and holographic methods, that seek to describe the strong interaction. Experimentally, the decay constants of charmonium and bottomonium can be extracted from the heavy vector to dileptons ($V\to l\,\bar{l}$) decay width. These constants display a monotonic decrease with increasing excitation level \cite{Navas:2024}.

This trend indicates that higher excited states are progressively less stable \cite{Braga:2018fyc, MartinContreras:2023oqs}, a defining feature of heavy quarkonium spectroscopy. In contrast, the decay constants of light mesons, such as the $\rho$, $\phi$, and $\omega$ mesons, are only known experimentally for their ground states \cite{Navas:2024}. However, from theoretical approaches such as Relativistic Hamiltonian Dynamics \cite{Krutov:2016uhy}, Light-Front Dynamics \cite{Aliev:2009gj, Vary:2018pmv}, or the Nambu-Jona-Lasinio Model \cite{Carrillo-Serrano:2015uca}, one can infer the decay constants of excited states. These theoretical estimates suggest a decrease in decay constants with increasing radial excitation level, suggesting that this pattern may be universal in meson spectroscopy. It should be emphasized that, for heavy–light mesons such as D and B, the decay constants are extracted from electroweak processes rather than electromagnetic ones. Consequently, the underlying mechanism differs from that of heavy quarkonia, where decay constants are obtained from electromagnetic decays into dileptons.

However, existing holographic models still have limitations in describing heavy quarkonium spectroscopy, as they cannot simultaneously reproduce the mass spectrum and decay constants with high accuracy. For example, Ref.~\cite{Grigoryan:2010pj} improved the masses and decay constants of $J/\psi$ and $\psi'$ by modifying the holographic potential, but failed to reproduce the experimental observation that the decay constants decrease with increasing excitation level. Refs.~\cite{Evans:2006ea, Afonin:2011ff, Afonin:2012jn, Braga:2015jca} introduced a finite $z_{uv}$ cutoff in the AdS slice to set the ultraviolet scale, which led to a decreasing trend of decay constants with higher excitation levels. However, the slope of this decrease was much smaller than that observed in the experimental results, providing only qualitative agreement. Another class of holographic models proposed in Ref.~\cite{Braga:2018hjt} successfully addressed the \emph{slope problem} of the decay constants and accurately reproduced the spectra of charmonium and bottomonium decay constants. However, they could not simultaneously provide a precise description of the mass spectrum. In Refs. \cite{MartinContreras:2019kah, MartinContreras:2021bis}, the authors inferred the role played by the ultraviolet (UV) limit of the dilaton field $\Phi(z)$ in determining decay constants. Therefore, modifying this limit allows us to induce the expected decrease in decay constants. Thus, transformations like \emph{isospectrality} \cite{MartinContreras:2023eft} or geometric deformations (as UV cutoffs) also modify decay constants.  

Recently, machine learning techniques, particularly multilayer perceptrons (MLPs), have shown significant promise in addressing complex scientific challenges \cite{Ma:2023ml, Zhou:2023fair, Pang:2024ijmpe, Li:2023pnn, Pu:2023dl, Pang:2021npa, Steinheimer:2019iso, Wang:2023lightning, Huang:2025ml, He:2023ml, He:2021ml}. The universal approximation theorem \cite{Hornik:1989} establishes that an MLP can approximate any continuous function, provided a sufficient number of hidden neurons are present. This capability renders MLPs highly effective for solving partial differential equations \cite{Raissi:2019pinn, Soma:2023eos, Karniadakis:2021piml, Shi:2022hq} and modeling inverse problems. Compared to traditional polynomial fitting or parametrization approaches, MLPs offer greater expressive power and enhance predictive accuracy while maintaining physical constraints. Beyond MLPs, other deep learning architectures such as Transformers have also been successfully applied to hadron physics~\cite{Tong:2026MesonNN} and has also been employed to extract the effective masses of chromo-electric and chromo-magnetic gluons from lattice QCD thermodynamic data \cite{Mei:2025}.

To the best of our knowledge, this work represents the first application of MLPs to reconstruct the dilaton field directly $\Phi(z)$ within a bottom-up AdS/QCD model. Unlike previous approaches that rely on \emph{ad hoc} parametrizations, our method uses experimental data on heavy quarkonium masses and decay constants to determine $\Phi(z)$  in a data-driven manner. This allows for a unified and highly accurate description of both spectroscopic and decay properties, addressing a longstanding challenge in holographic QCD.

In recent years, the rapid advancement of machine learning techniques \cite{He:2023mlnuclear, Zhou:2018eos, Pang:2019eos, Wang:2022piml, Zhao:2022bnn, Du:2020dljet, Du:2022jettomo, Jiang:2021eos, Shi:2022lattice, Zeng:2025tcz} has facilitated the introduction of deep learning into the study of holographic QCD, offering novel perspectives for model construction \cite{Akutagawa:2020dladsqcd, Hashimoto:2018dlholo, Yan:2020dlbh, Hashimoto:2022dilaton, Song:2021adsdbm, Chang:2024hqanisotropy, Ahn:2024dlconductivity, Gu:2024dlee, Li:2023bhmetric, Cai:2024qcdml, Ahn:2024bhrecon}. In contrast to traditional bottom-up models that depend on \emph{ad hoc} specified potentials or dilaton profiles, data-driven approaches leverage experimental or lattice QCD data to determine model parameters, thereby enhancing both reliability and interpretability. Recent studies demonstrate that holographic models integrated with neural networks can effectively address inverse problems and achieve superior performance in the non-perturbative regime of the strong interaction \cite{Hashimoto:2018dlholo, Akutagawa:2020dladsqcd, Ahn:2024dlconductivity}. In this manuscript, deep learning techniques were employed to reconstruct the dilaton field using an MLP. The reconstructed dilaton not only improves the fitting accuracy but also reveals clear physical mechanisms: in the ultraviolet region, the non-quadratic corrections through the background function dominate the decrease of the decay constants, while in the infrared region, the accelerated growth enhances the holographic potential $U(z)$ and produces nonlinear radial Regge trajectories.The model predictions show excellent agreement with experimental data, with root-mean-square errors of approximately 1.26\% for charmonium and 3.32\% for bottomonium, thereby providing an effective tool for investigating heavy-quarkonium spectroscopy at zero temperature.

The structure of this paper is as follows. Section \ref{0} introduces the motivation and construction of the holographic model, with particular emphasis on reconstructing the dilaton field using an MPL. Section \ref{1} presents the spectroscopy of charmonium and bottomonium at zero temperature, including both mass spectra and decay constants, and systematically compares predictions with experimental data to assess the model validity. Section \ref{5} summarizes the main results and innovations of this work and discusses potential applications of the model in future studies of heavy quarkonia.
  
\section{Holographic Model}
\label{0}
Reference \cite{MartinContreras:2021bis} proposed a forward-direction construction within the bottom-up AdS/QCD framework, introducing an appropriate dilaton profile to simultaneously describe the mass spectra and decay constants of heavy quarkonia. Although the model is phenomenological, it demonstrates strong agreement with experimental data. This approach has also been applied and refined in various holographic models to investigate the spectral properties of heavy quarkonia at finite temperature.  Building on this forward framework, the corresponding vector-field action and equations of motion are presented below as the foundation for the subsequent inverse analysis.

At zero temperature, the background geometry is taken to be the five-dimensional anti–de Sitter (AdS$_5$) spacetime, whose metric in the Poincar\'{e} patch is

\begin{equation}
ds^2 = \frac{R^2}{z^2} \left( \eta_{\mu\nu} \, dx^\mu dx^\nu - dz^2 \right) ,
\end{equation}

\noindent where $z$ is the holographic coordinate,  with $z \to 0$ corresponding to the ultraviolet (UV) boundary and $z \to \infty$ to the infrared (IR) region. Here $\eta_{\mu\nu} = \mathrm{diag}(1,-1,-1,-1)$ is the four-dimensional Minkowski metric, and $R$ is the AdS radius.

The action for the gauge field is written as

\begin{equation}
S_{\text{Vector}\, Q\bar{Q}} = -\frac{1}{4g_5^2} \int d^5x \, \sqrt{-g} \, e^{-\Phi(z)} \, g^{mp} g^{nr} F_{mn} F_{pr} ,
\end{equation}

\noindent where $F_{mn} = \partial_m V_n - \partial_n V_m$ is the field strength tensor of the five-dimensional gauge field $V_m$; $g$ denotes the determinant of the five-dimensional metric; $g^{mp}$ and $g^{nr}$ are the inverse metric components used to raise covariant indices to contravariant ones and perform contractions; and $\Phi(z)$ is the dilaton field, whose exponential factor $e^{-\Phi(z)}$ encodes the nonperturbative effects of QCD.

By varying the action with respect to $V_m$, the equation of motion is obtained as

\begin{equation}
\partial_m\left(\sqrt{-g} \, e^{-\Phi(z)} \, g^{np} g^{mr} F_{pr}\right) = 0 ,
\end{equation}

\noindent  where $F_{mn} \equiv \partial_m V_n - \partial_n V_m$.

In the bottom-up AdS/QCD framework, heavy vector quarkonia can be described by a five-dimensional Abelian massless gauge field. This setting follows from the standard field–operator correspondence \cite{Aharony:1999ti}, where the bulk mass is related to the scaling dimension $\Delta$ of the boundary operator creating hadrons by

\begin{equation}
M_5^2 R^2 = (\Delta - 1)(\Delta -3) ,
\end{equation}

\noindent and for the vector current $\bar{Q}\gamma_\mu\,Q$ operator with $\Delta = 3$, one obtains

\begin{equation}
M_5^2 R^2 = 0.
\end{equation}

In the axial gauge $V_z = 0$ and with the transverse condition $\partial^\mu V_\mu = 0$, we take the plane-wave ansatz

\begin{equation}
V_\mu(x,z) = \epsilon_\mu \, v(z) \, e^{ip\cdot x} ,
\end{equation}

\noindent where $\epsilon_\mu$ is the polarization vector and $p^2 = m_n^2$ is the on-shell mass condition for vector mesons. Substituting this ansatz into the equation of motion leads to the \emph{Sturm-Liouville} equation for the vector bulk modes

\begin{equation}\label{S-L-eqn}
\partial_z \left( e^{-B(z)} \partial_z v_n(z) \right) + m_n^2 e^{-B(z)} v_n(z) = 0 ,
\end{equation}

\noindent where the \emph{background function} $B(z)$ is defined as

\begin{equation}
B(z) = \Phi(z) - \log\left(\frac{R}{z}\right).
\end{equation}

One of the main consequences of confinement is the emergence of bound states of quarks, i.e., hadrons. At the holographic level, confinement is achieved by breaking bulk conformal invariance.  To do so, it is customary to use a hard cutoff in AdS space \cite{Boschi-Filho:2002xih, Erlich:2005qh} (hard-wall model) or a dilaton field $\Phi(z)$, static \cite{Karch:2006pv} or dynamically \cite{Batell:2008zm} generated soft-wall model.  Analogously to quantum mechanics, the hard-wall model is like solving the square-well potential, and the soft-wall model is like solving a 2-dimensional harmonic oscillator. It is in this context that \emph{in bottom-up models, confinement is realized with the emergence of bound states in the AdS bulk, dual to hadrons at the boundary.} 

In the particular case of the soft-wall model, the dilaton is defined by a static quadratic form $\Phi(z)=\kappa^2\,z^2$, with $\kappa$ the energy scale that sets hadron masses. The emerging radial Regge trajectories $M_n^2$ are linear, and the decay constant spectrum $f_n$ is degenerate: 

\begin{eqnarray}
    M_n^2&=&4\kappa^2(n+1),\\
    f_n^2&=&\frac{2\kappa^2}{g_5^2},
\end{eqnarray}

\noindent where $n$ is the radial excitation number. 

However, the linearity observed in the mass spectrum does not satisfy the phenomenological constraints imposed by the Bethe-Salpeter theory \cite{Chen:2018bbr}, since the heavy quark mass becomes relevant in this case. On the pure phenomenological side, from non-relativistic QCD, the one-gluon exchange term, which is Coulomb-like, has a stronger \emph{inertial} effect on the heavy meson ground state than the linear confinement term \cite{Lucha:1996ax}.  Thus, for heavy vector mesons, a \emph{non-quadratic dilaton is expected} \cite{MartinContreras:2025wnh}. When high radial excitations are considered, the confinement term in the Cornell potential becomes relevant, still non-linear, since in this limit the mass squared grows as $n^{2/3}$.  Linear Regge trajectories observed in light meson systems arise from relativistic effects. Therefore, it is consistent to expect a non-linear parametrization for Regge trajectories.  

Regarding the decay constants for vector mesons, neither the hard-wall nor the soft-wall model describes them properly. In the former case,  decay constants increase with radial excitations \cite{Polchinski:2002jw}. In the latter, the decay constants are degenerate. Although this outcome may seem counterintuitive, it is expected in the context of large-$N_c$ QCD. At large $Q^2$, the asymptotics of the vector meson correlation functions are given by \cite{Wise:1997sg, Afonin:2007gd}

\begin{equation}
\Pi(Q^2) \sim \frac{N_c}{12\pi^2} \left[ \ln\left(\frac{Q^2}{m_Q^2}\right) - \frac{6\,m_Q^2}{Q^2} + \mathcal{O}\left(\frac{m_Q^4}{Q^4}\right) \right]\sim \sum_n\frac{f_n^2}{Q^2-M_n^2}
\end{equation}

When the sum is approximated using the Euler-Maclaurin summation formula \cite{Afonin:2004yb}, the decay constants, as residues of $\Pi(q^2)$, can be written as

\begin{equation}
    f_n^2\propto \frac{d\,M_n^2}{d\,n}\propto \text{constant},
\end{equation}

\noindent which holds at least for $n\to\infty$ and linear $M_n^2$. This picture is consistent with the \emph{string picture of mesons}. Notice that, since the meson size is given by the string length $l\sim M_n$, for the leptonic decay width one gets $\Gamma_{V_n\to e^+e^-}\sim1/l\sim1/M_n$. However, on the other hand, $\Gamma_{V_n\to e^+e^-}\sim f_n^2/M_n$. Thus, $f_n\sim$ constant, in agreement with the soft-wall holographic counterpart. 

Nevertheless, this particular scaling behavior is expected for light mesons, for which the relativistic string picture applies. In the heavy case, the non-relativistic kinematics, the mass dominance $M_n\sim 2\,m_Q$, different size scaling (heavy mesons do not scale linearly with the mass), and the wave-function suppression at the origin (as described by the Van Royen–Weisskopf formula), break this ideal picture \cite{Lucha:1991vn}. 

Therefore, this phenomenological analysis motivates moving beyond the quadratic static dilaton ansatz to describe both the electromagnetic decay constants and the radial Regge trajectories of heavy vector mesons.

In this work, we do not assume an explicit analytic form for $\Phi(z)$; instead, we determine it through an inverse method based on experimental data, ensuring that the resulting $\Phi(z)$ satisfies both IR and UV constraints for the potential defined as 

\begin{eqnarray} \label{cond1}
    \left.U(z)\right|_{z\to0}&=&\frac{A}{z^2}\\ \label{cond2}
    \left.U(z)\right|_{z\to\infty}&=&B\,z^2+f(z)
\end{eqnarray}

\noindent where the condition \eqref{cond1} corresponds to the UV constraint, defined by the field/operator itself, since the parameter $A$ carries the information about the conformal dimension $\Delta$. In bottom-up models, this parameter is fixed to be $3/4$ for vector states. The second condition \eqref{cond2} controls variations from the standard quadratic dilaton form observed in the Soft-wall model. Any non-linearities expected in the spectrum will be encoded in the function $f(z)$. These conditions provide a suitable background for subsequent calculations of the mass spectrum and decay constants.

By redefining the bulk field with the \emph{Bogoliubov transformation} as $v_n(z) = e^{B(z)/2} \psi_n(z)$ using the background function, the Sturm-Liouville equation \eqref{S-L-eqn} can be cast into a \emph{Schrödinger-like} form

\begin{equation}\label{sch-eq}
- \psi_n''(z) + U(z)\, \psi(z) = m_n^2 \,\psi(z), 
\end{equation}

\noindent where the holographic bottom-up confining potential for vector mesons is written as:

\begin{equation}
    U(z)=\frac{1}{4}B'(z)^2-\frac{1}{2}B''(z).
\end{equation}

For numerical implementation, the effective potential function \( U(z) \) is directly expanded in terms of the derivatives of the dilaton field \( \Phi(z) \) as follows:

\begin{equation}\label{con-pot}
U(z) = \frac{3}{4z^2} + \frac{\Phi'(z)}{2z} + \frac{\Phi'(z)^2}{4} - \frac{\Phi''(z)}{2}
\end{equation}

In the framework of holographic QCD, the decay constant $f_n$ of a vector meson is determined by the behavior of the five-dimensional gauge field near the UV boundary. For the $n$-th eigenstate, with mass denoted by $M_n$ and radial wave-function $v_n(z)$, one has

\begin{equation}\label{decay}
f_n = \frac{1}{g_5\, M_n} \lim_{z \to 0} \left[ e^{-B(z)}\, \partial_z v_n(z) \right],
\end{equation}

\noindent where $g_5$ is the five-dimensional gauge coupling constant and $B(z)$ is the background function defined earlier. The limit $z \to 0$ corresponds to the UV boundary of the AdS space, where the five-dimensional field is matched to the external source of QCD in four dimensions, thereby relating the boundary current to the four-dimensional vector current matrix element by the holographic dictionary.

\section{Modeling Holographic QCD with MLP}
\label{1}
\begin{figure}
    \centering
    \includegraphics[width=0.8\textwidth]{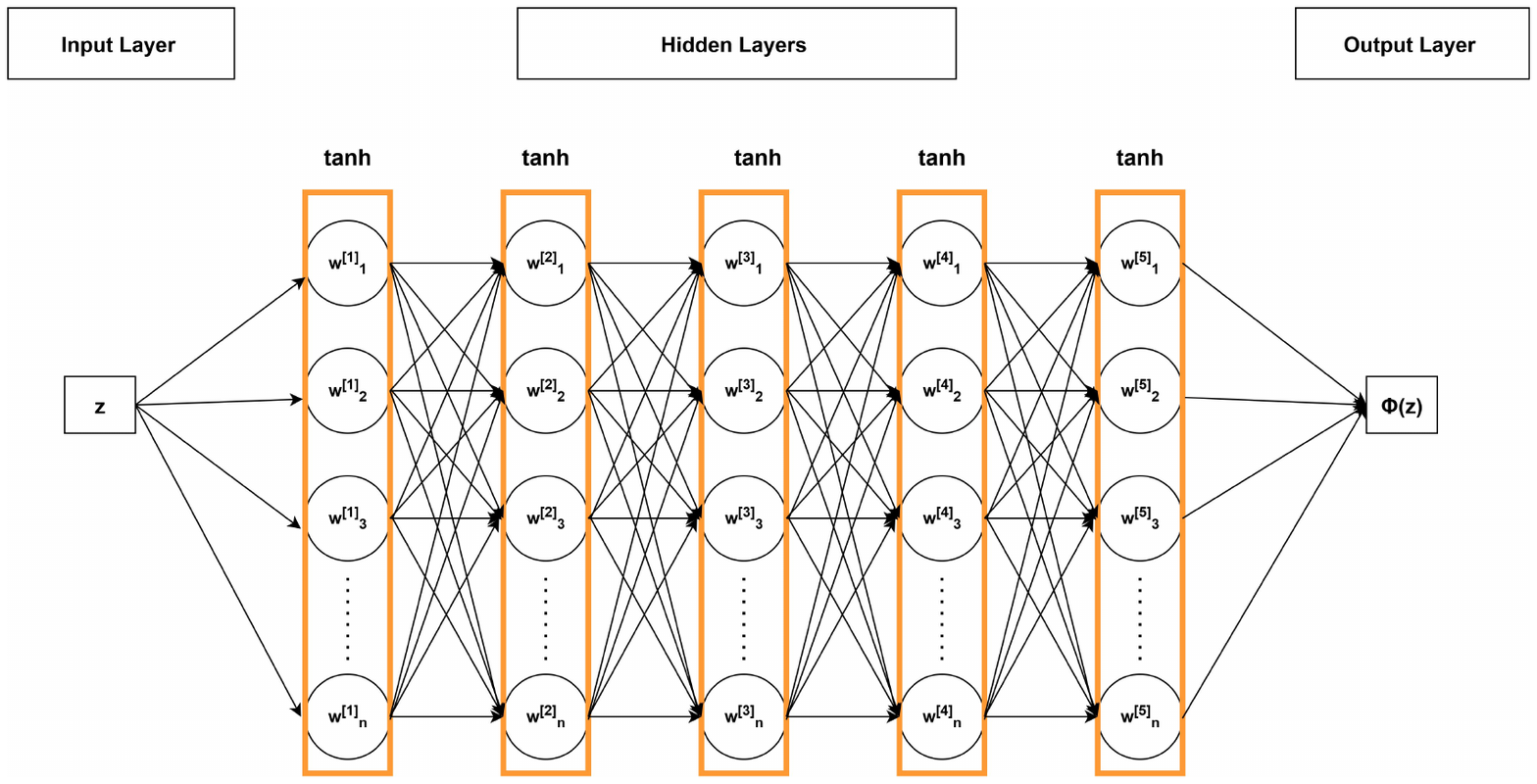}
    \captionsetup{aboveskip=-10pt,belowskip=0pt}
    \caption{A neural network flowchart for reconstructing the dilaton field $\Phi(z)$.}
    \label{fig:1}
\end{figure}

Based on the MLPs, we reconstruct the dilaton field $\Phi(z)$ within the AdS/QCD framework. The background profile is directly inferred from experimental data, specifically the mass spectrum and decay constants of charmonium and bottomonium states. The architecture of the neural network is illustrated in Fig.~\ref{fig:1}, consisting of five hidden layers with 128 neurons each. Trainable parameters include the weights and biases of all hidden layers, denoted as $w^{(1)}_n, w^{(2)}_n, \dots, w^{(5)}_n$, as well as the output layer parameters $v_n$, all optimized using the Adam gradient descent algorithm. The training objective is to minimize the mean squared error between the predicted mass spectrum and the experimental data. To provide a unified quantitative measure of the overall fitting accuracy, we introduce the root-mean-square (RMS) deviation, defined as

\begin{equation}
\label{eq:RMS}
\mathrm{RMS}
= \sqrt{\frac{1}{N}\sum_{i=1}^{N}\left(\frac{\mathcal{O}_i^{\mathrm{pre}}-\mathcal{O}_i^{\mathrm{exp}}}{\mathcal{O}_i^{\mathrm{exp}}}\right)^2}\times 100\%,
\end{equation}

\noindent where \(N\) denotes the total number of observables, including both masses and decay constants, and \(\mathcal{O}_i^{\mathrm{pre}}\) and \(\mathcal{O}_i^{\mathrm{exp}}\) represent the predicted and experimental values of the \(i\)-th observable, respectively. This definition provides a single indicator that simultaneously reflects the model performance across both spectroscopic and dynamical quantities, offering a clear, intuitive measure of overall agreement with experimental data.

The key idea is as follows: the function $\Phi'(z)$ is represented by a deep neural network that takes the holographic coordinate $z$ as input. During training, the network outputs $\Phi'(z)$, and its exact derivative $d\Phi'(z)/dz$ is computed by automatic differentiation. These quantities are combined according to Eq. \eqref{con-pot} to reconstruct the potential $U(z)$. The loss function, defined as the \textbf{SmoothL1Loss} between this reconstructed potential and the target potential from a pre-trained model, is minimized using the \emph{Adam optimizer} with a cosine-annealing learning rate schedule. Training converges within $10^4$ iterations, with automatic gradient clipping preventing instability. The entire procedure operates in a mesh-free manner without explicit discretization or Hamiltonian matrix construction.

A $\tanh$ activation function follows each hidden layer, while the output layer uses no activation. To enforce ultraviolet (UV) boundary regularity, the network output is structurally transformed to satisfy the condition $\Phi(0) = 0$, thereby ensuring consistency with the theoretical requirements of the AdS/QCD model.

Once the dilaton field is obtained, its first and second derivatives are computed by automatic differentiation and substituted into Eq.~\eqref{con-pot} to construct the effective potential $U(z)$. The Schrödinger-like equation, Eq.~\eqref{sch-eq}, is discretized on a finite grid and solved by Hamiltonian diagonalization to extract the mass eigenvalues $m_n$ and corresponding wave functions $\psi_n(z)$. The decay constants $f_n$ are computed using Eq.~\eqref{decay}, which depends on the near-boundary behavior of both $\Phi(z)$ and $\psi_n(z)$. 

The reconstructed dilaton field exhibits a non-quadratic expansion near the UV boundary, while its rapid growth in the IR reflects the confining nature of QCD. This non-quadratic IR behavior is crucial for reproducing the non-linear Regge trajectories and decreasing decay constants observed in heavy quarkonia. The deviation from a purely quadratic form underscores the importance of heavy-quark mass effects, which break the relativistic string picture applicable to light mesons.

\begin{figure}
    \centering
    \captionsetup{justification=raggedright,singlelinecheck=false}
      \includegraphics[scale=0.5]{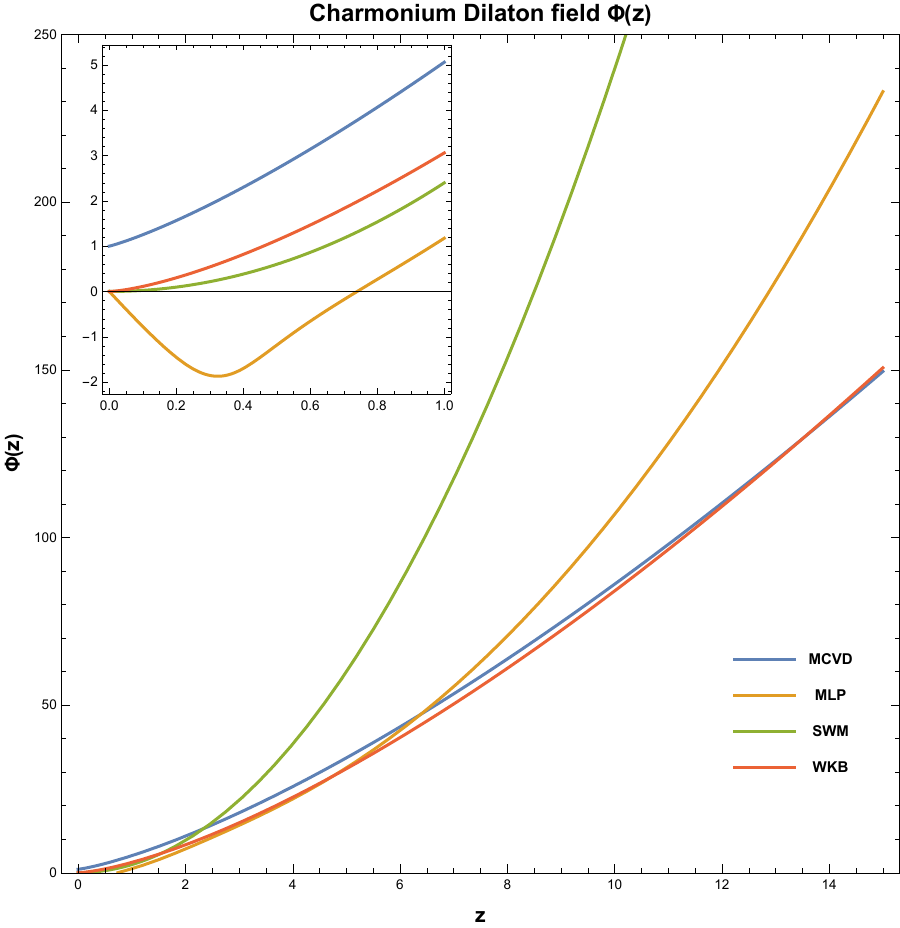}
		 \includegraphics[scale=0.5]{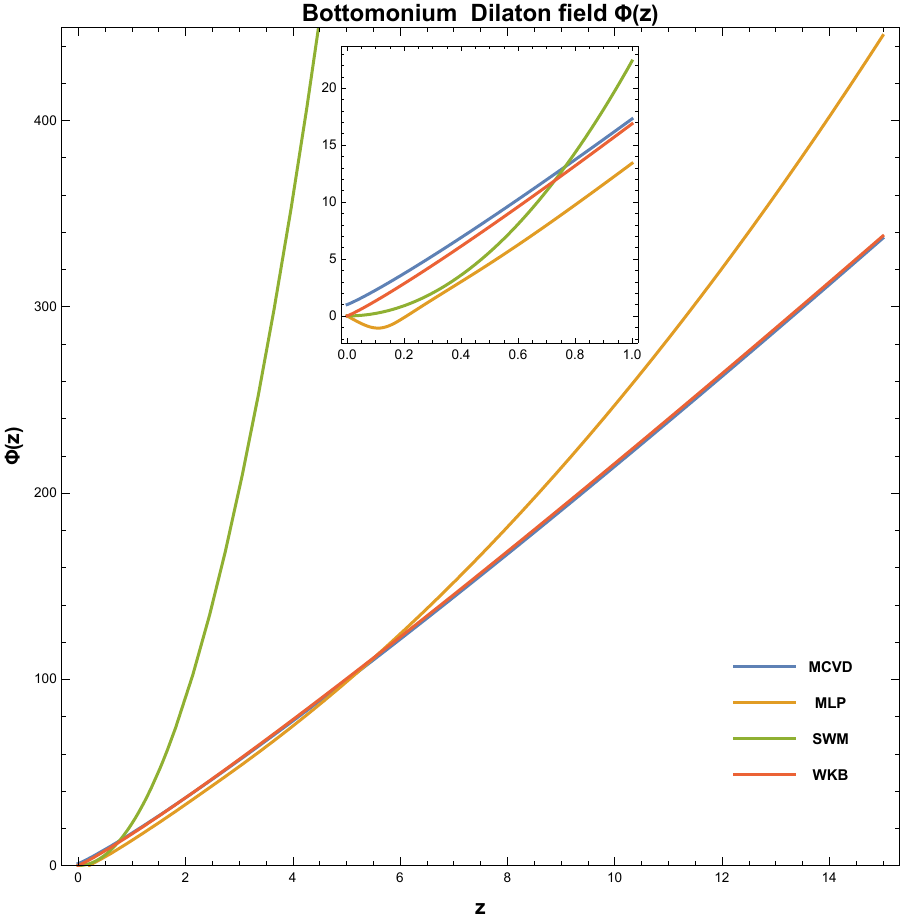}
    \captionsetup{aboveskip=0pt,belowskip=0pt}
	\captionsetup{justification=raggedright,singlelinecheck=false}
    \caption{Dilaton fields $\Phi(z)$ for charmonium (left) and bottomonium (right) obtained from different approaches: Non-linear approach MCVD \cite{MartinContreras:2021bis} (blue), Machine Learning Perceptron (MLP) in orange, soft-Wall model \cite{Karch:2006pv}, and WKB approach \cite{MartinContreras:2025wnh}.}
	\label{fig:2}   
\end{figure}

In the final implementation, we adopted a more rigorous training strategy. Instead of directly fitting the effective potential $U(z)$, we employed a neural network to reconstruct the dilaton field $\Phi(z)$, with the ultraviolet boundary condition $\Phi(0)=0$ explicitly imposed to ensure compatibility with the AdS/QCD framework. Once the dilaton profile was obtained, the effective potential $U(z)$ was analytically derived from $\Phi(z)$ and its first and second derivatives. This procedure guarantees that the definition of the potential strictly follows the holographic construction, while avoiding the introduction of an \emph{ad hoc} base-plus-correction structure. As a result, the model successfully reproduces the decreasing trend in decay constants and predicts the masses of higher-excited states. (Detailed code: \url{https://github.com/yu11c/DilatonNN}).

After the training converges, the reconstructed dilaton field is shown in Fig.~\ref{fig:2}. In the UV region, $\Phi(z)$ exhibits an approximate non-quadratic expansion, different from the expected quadratic one, characteristic of the soft-wall model, lifting the degeneracy of the decay constants. In contrast, in the IR region, it grows rapidly, reflecting the confinement effect. These corrections are essential for reproducing the experimental spectrum. Compared to the conventional analytical soft-wall ansatz, our data-driven reconstruction yields IR growth that is more consistent with experimental requirements, thereby improving the description of higher-excited states.

To further highlight the advantages of this work, Fig.~\ref{fig:2} presents a comparison of the MLP reconstruction results for the dilaton $\Phi(z)$ with three existing models: the traditional soft-wall model \cite{Karch:2006pv}, the WKB analysis \cite{MartinContreras:2025wnh}, and the results for a non-quadratic dilaton \cite{MartinContreras:2021bis}, which we call MCVD (Martin-Contreras, Vega, and Diles). 

In the soft-wall model, the quadratic dilaton leads to degenerate decay constants, tracing back the phenomenology expected from the Large-$N$ limit, inherent to this model. In the UV region, the WKB and soft-wall model dilatons behave similarly. However, they differ in the infrared limit, since they reproduce different concavities in the Regge trajectories: the WKB is intended for non-linear whilst the soft-wall reproduces linear ones. The MCVD dilaton imposes a different UV behavior to break the soft-wall model degeneracy \cite{Braga:2018hjt}, but conserves the non-quadratic behavior required for heavy quarkonium spectrum. The dilaton reconstructed by the MLP naturally exhibits a non-quadratic expansion in a data-driven manner, and through the background function $B(z)=\Phi(z)-\log(z/R)$, it breaks the degeneracy of $f_n$ and reproduces the experimentally observed monotonic decrease of decay constants with radial excitation.
In contrast, the MLP reconstruction results are more consistent with the experimental data overall, particularly in reproducing both the mass spectrum and the decay constants with higher accuracy. This comparison further confirms the effectiveness of data-driven approaches in the study of nonperturbative QCD.

\begin{figure}
	\centering
    \captionsetup{justification=raggedright,singlelinecheck=false}
      \includegraphics[scale=0.5]{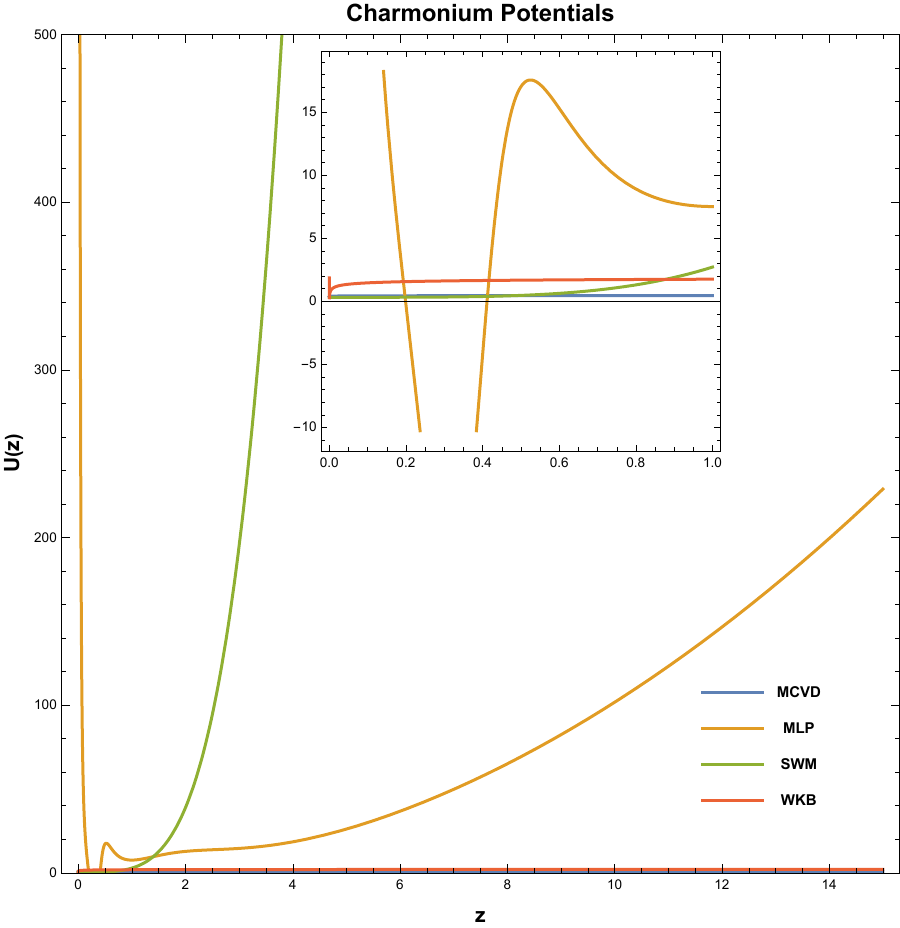}
		 \includegraphics[scale=0.5]{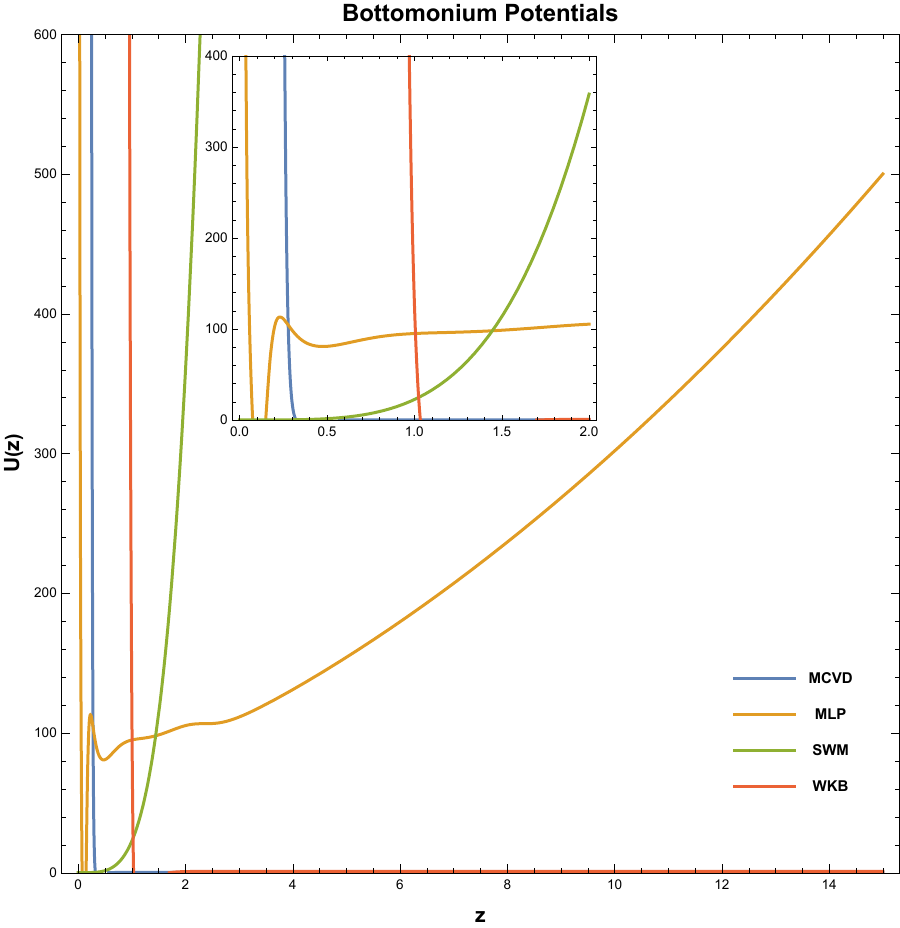}
	\caption{Potentials $U(z)$ for charmonium (left) and bottomonium (right) obtained from different approaches: Non-linear approach MCVD \cite{MartinContreras:2021bis} (blue), Machine Learning Perceptron (MLP) in orange, soft-wall model \cite{Karch:2006pv}, and WKB approach \cite{MartinContreras:2023eft}. }
	\label{fig:3}
\end{figure}

For completeness, Fig.~\ref{fig:3} presents the effective potentials $U(z)$ reconstructed for charmonium and bottomonium. In the ultraviolet region, the potentials exhibit near-harmonic-oscillator behavior. In contrast, in the infrared, they develop nonlinear corrections that are crucial for reproducing the experimental mass spectra and decay constants. The comparison highlights that the neural network reconstruction not only captures the dilaton profile but also provides a consistent effective potential, thereby reinforcing the reliability of the approach.

\vspace{0.5cm}  

\begin{table}[htbp]
\centering
\captionsetup{justification=raggedright,singlelinecheck=false}
\renewcommand{\arraystretch}{0.6} 
\begin{tabular}{lccccccc} 
\toprule 
\multicolumn{7}{c}{Charmonium States \( I^G(J^{PC}) = 0^+(1^{--}) \)} \\
\midrule 
\midrule 
\( n \) & State & \( M_{\text{Exp}} \) (MeV) & \( M_{\text{MLP}} \) (MeV) & \( \%M \) & \( f_{\text{Exp}} \) (MeV) &\( f_{\text{MLP}} \) (MeV) &  \( \%f \) \\
\midrule 
1 & \( J/\psi \) & \( 3096.900 \pm 0.006 \) & 3097.2 & 0.01 & \( 416.16 \pm 5.25\) & 416.2 & 0.01  \\
\midrule 
2 & \( \psi(2S) \) & \( 3686.097 \pm 0.010 \) & 3686.5 & 0.01 & \( 296.08 \pm 2.51\) & 296.1 & 0.01 \\
\midrule 
3 & \( \psi(4040) \) & \( 4040 \pm 4 \) & 4040.4 & 0.01 & \( 187.13 \pm 7.61\) & 186.9& 0.1 \\
\midrule 
4 & \( \psi(4415) \) & \( 4415 \pm 5 \) & 4416.5 & 0.03 & \( 160.78 \pm 9.70\) & 157.6 & 2.0\\
\bottomrule 
\end{tabular}
\vspace{0.1em}
\caption{Summary of results for charmonium states. Theoretical predictions based on neural network reconstruction results are listed in the $M_{\text{MLP}}$ and $f_{\text{MLP}}$ columns. The corresponding relative errors are shown in the $\%M$ and $\%f$ columns. Experimental data are taken from PDG~\cite{ParticleDataGroup:2024} and the total root-mean-square (RMS) error is $\delta_{\mathrm{RMS}} = 1.26\%$. }
\label{tab:charmonium}
\end{table}


\begin{table}[htbp]
\centering
\captionsetup{justification=raggedright,singlelinecheck=false}
\renewcommand{\arraystretch}{0.6} 
\begin{tabular}{lccccccc}
\toprule
\multicolumn{8}{c}{Bottomonium States \( I^G(J^{PC}) = 0^+(1^{--}) \)} \\
\midrule
\midrule
\( n \) & State & \( M_{\text{Exp}} \) (MeV) & \( M_{\text{MLP}} \) (MeV) & \( \%M \) & \( f_{\text{Exp}} \) (MeV) &\( f_{\text{MLP}} \) (MeV) &  \( \%f \) \\
\midrule
1 & \( \Upsilon(1S) \) & \( 9460.4 \pm 0.1 \) & 9460.42 & 0.0002 & \( 714.99 \pm 2.40 \)&\( 715.1 \) & 0.01 \\
\midrule
2 & \( \Upsilon(2S) \) & \( 10023.4 \pm 0.5 \) & 10023.42 & 0.0002 & \( 497.37 \pm 2.23 \) & \( 497.4 \) &0.009 \\
\midrule
3 & \( \Upsilon(3S) \) & \( 10355.1 \pm 0.5 \) & 10355.08 & 0.0002 & \( 430.11 \pm 1.94\) &\( 430.2 \) & 0.2 \\
\midrule
4 & \( \Upsilon(4S) \) & \( 10579.4 \pm 1.2 \) & 10579.41 & 0.0001 & \(340.65 \pm 9.08 \) &\( 340.9 \) & 0.4\\
\midrule
5 & \( \Upsilon(10860) \) & \( 10885.2^{+2.6}_{-1.6}  \) & 10863.27 & 0.2 &\( 310.91 \pm 60.80 \) & \( 316.66 \) & \( 1.8 \)\\
\midrule
6 & \( \Upsilon(11020) \) & \( 11000 \pm 4 \) & 11149.99 & 1.3 &\(240.13 \pm 27.70 \) & \( 267.33 \) & \( 11.3 \)\\
\bottomrule
\end{tabular}
\vspace{0.1em}
\caption{Summary of results for bottomonium states. Theoretical predictions based on neural network reconstruction results are listed in the $M_{\text{MLP}}$ and $f_{\text{MLP}}$ columns. The corresponding relative errors are shown in the $\%M$ and $\%f$ columns. Experimental data are taken from PDG \cite{ParticleDataGroup:2024} and the total root-mean-square (RMS) error is $\delta_{\mathrm{RMS}} = 3.32\%$. }
\label{tab:bottomonium}
\end{table}
\FloatBarrier   


\begin{figure}
	\centering
    \captionsetup{justification=raggedright,singlelinecheck=false}
      \includegraphics[scale=0.5]{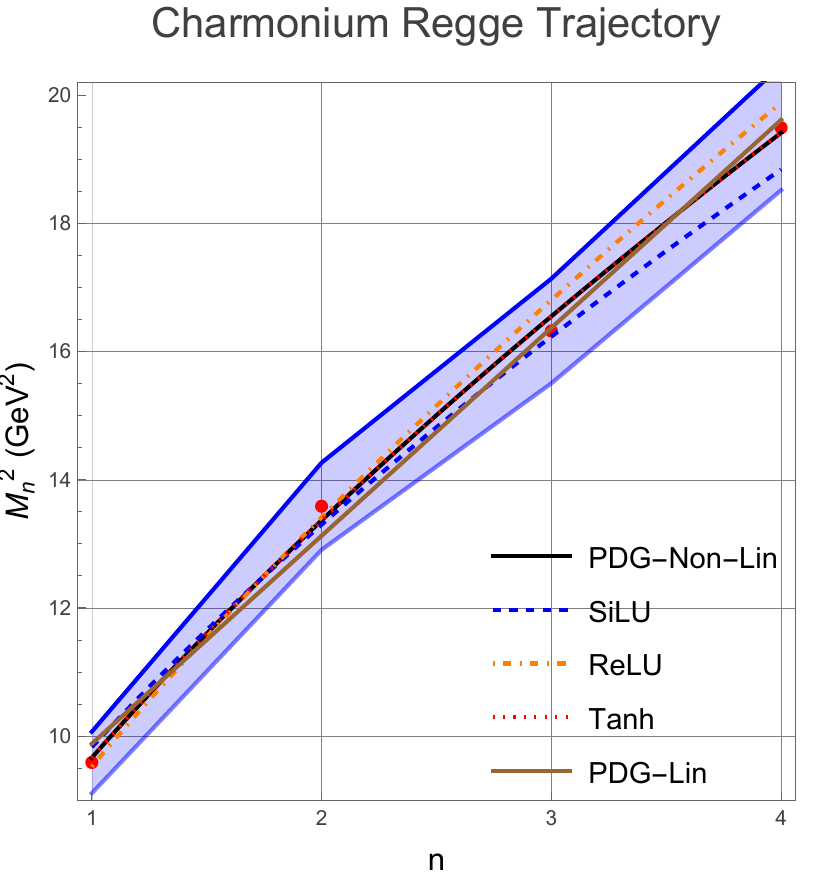}
		 \includegraphics[scale=0.51]{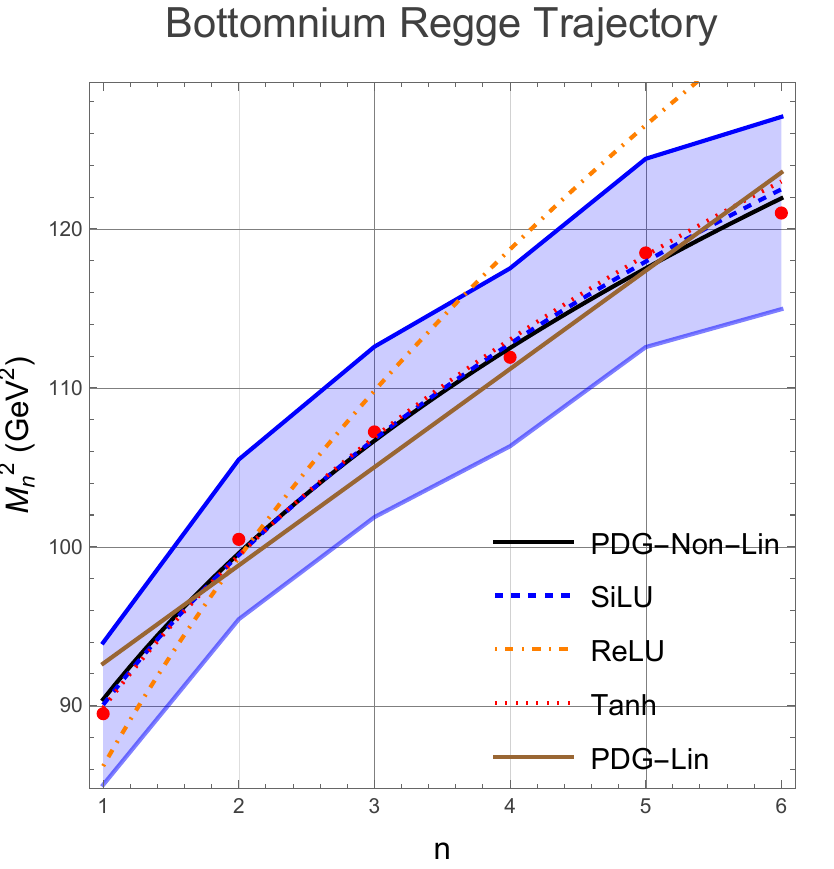}
	\caption{Chew-Frautschi plot for radial non-linear Regge trajectories for charmonium (left) and bottomonium states (right) adjusted using different activation functions: Tanh (red dotted), ReLU (orange dot-dashed), and SiLU (blue dashed). The purple-shaded region represents an \emph{ad hoc} error band corresponding to a relative error of 5\% for the experimental data. The experimental fit, based on data from PDG \cite{Navas:2024}, is shown as the black line. For reference, we plot the experimental linear fit $M_n^2=a(n+b)$ in brown to make contrast with the non-linear parametrization.}
	\label{fig:4}
\end{figure}

In the numerical implementation, we adopted a division of the dataset into training and prediction subsets in order to examine the generalization capability of the neural network approach. Specifically, in the charmonium sector, the first three experimental states were used as training data, and the trained network was subsequently employed to predict the mass and decay constant of the fourth state. In the bottomonium sector, the first four experimental states were included in the training set, while the trained network was further applied to predict the properties of the two higher excitations. This partitioning strategy demonstrates that the constructed neural network not only achieves accurate fits to the available experimental data but also provides reasonable predictions for higher excited states, thereby highlighting the robustness and reliability of the method in heavy quarkonium spectroscopy.

From Tables~\ref{tab:charmonium} and~\ref{tab:bottomonium}, the predicted mass spectrum and decay constants for the ground state and the first excited state are in excellent agreement with the experimental values, with relative errors remaining within acceptable ranges. As the excitation level increases, the deviations become larger, but the overall trend remains consistent with the experimental spectrum. These results indicate that the neural network not only satisfies the boundary condition $\Phi(0)=0$, but also effectively captures the systematic behavior of the spectrum. The comparison in the tables further supports the model's validity and verifies the neural network's reliability in reconstructing the dilaton field.

To test the accuracy and efficiency of the reconstruction  mechanism using data-driven techniques, we change the activation function while keeping the epochs, neurons, and initialization strategy identical. This kind of test is relevant since the dataset is small in size and some of the data lies close to the boundary of the input domain. To do so, we consider three different activation functions: Tanh, SiLU, and ReLU. As an accuracy test, we define a 5\% error band.  The results for this analysis are depicted in Fig. \ref{fig:4}.

Following Fig. \ref{fig:4}, all three activation functions interpolate the lower-$n$ points accurately, although the ReLU activation diverges for the bottomonium fit. The Tanh activation function yields a prediction that matches experimental data (PDG) within 0.5\% relative error, whereas ReLU and SiLU overstimate the mass for the higher excited states by more than 5\% and 8\%, respectively. This is not a statistical fluctuation. Recall that Tanh and SilU are bounded, while ReLU is piecewise-defined functions that grow unbounded, causing that near the domain boundary ($n=4$ for charmonium and $n=6$ for bottomonium, the fit develops higher masses. Therefore, this feature is a deterministic consequence of the mathematical properties of the activation functions and of how they implicitly regularise the network \cite{Xu2020sep}.

The origin of this uncontrolled extrapolation is understood by looking at how the activation function works. A  neural network with ReLU activation $\sigma(x) = \max(0,x)$ is piecewise linear and unbounded for $X>0$. For large inputs, a single ReLU unit behaves as a linear function whose slope is given by the incoming weight. As shown by Ziyin et al. \cite{ziyin2020aug} and Xu et al. \cite{Xu2020sep}, the extrapolation of a ReLU network outside the training domain is therefore a linear function with a slope that can be arbitrarily large. When only five points are available, the optimizer finds it advantageous to assign a large positive weight to a neuron that becomes active just beyond $n=4$. This allows the network to fit the interior points with a small curvature while satisfying the outermost point with a steep final segment, a configuration that minimizes the mean squared error on the four low-$n$ points at the expense of a large error on the boundary point. Because no penalty is placed on the magnitude of the weights, the network exhibits what we may call \emph{unrestrained extrapolation}.

In the case of SiLU activation, $\sigma(x) = x\cdot\text{sigmoid}(x)$, is asymptotically linear for $z\to+\infty$. Hence, it suffers from the same deficiency, albeit with a somewhat smoother transition around the origin. This fact is noticeable in the charmonium case.

In contrast, the hyperbolic tangent $\tanh(x) \in (-1,1)$ is bounded and saturating. For large $|x|$, its derivative decays as $4e^{-2|x|}$, so the output of a neuron cannot grow indefinitely. The decay of the derivative forces the function to flatten when the input moves far from the training data. The optimization procedure, driven solely by the empirical risk without explicit weight decay, must then strike a balance: increasing the outgoing weights to push the tail upward also shifts the entire function, including the interior points, thereby increasing the overall loss. This self–limiting mechanism acts as an \emph{implicit regulariser}, favoring solutions with moderate extrapolation slopes \cite{maennel2018aug}. Furthermore, Tanh is infinitely differentiable, which prevents the network from developing a kink at $n=4$ for charmonium and $n=6$ for bottomonium. The ReLU network can freely place a change of slope exactly at the penultimate training point; the Tanh activation function cannot, because any rapid change must be smoothed out over a finite interval, coupling the last two points and thereby transferring gradient information that defends the boundary point.

The main conclusion of this test is that, when reconstructing or fitting hadron spectra from very sparse data (e.g., five radial excitations), bounded and saturating activation functions such as Tanh effectively tame the extrapolation instability that plagues unbounded activations. In particular, Tanh consistently yields the lowest relative error at the largest quantum number, acting as an implicit regulariser that prevents the network from developing unrealistically steep slopes outside the training support. The unbounded functions ReLU and SiLU, by contrast, suffer from uncontrolled growth at the boundary, compromising the prediction of the highest state unless explicit regularisation (weight decay, dropout) is applied. This finding stresses that the activation function is not merely a technical hyperparameter but a decisive component of the model’s inductive bias, especially in data‑scarce hadron spectroscopy problems.

\section{Summary}
\label{5}
In this study, we employ an MLP to reconstruct the dilaton field $\Phi(z)$ within the AdS/QCD framework directly from experimental data. The neural network is used to parameterize $\Phi'(z)$, and its derivative $\Phi''(z)$ is obtained by automatic differentiation. These quantities are combined according to Eq.~\eqref{con-pot} to construct the effective potential $U(z)$.

The Schrödinger-like equation is discretized on a finite grid using a second–order finite-difference scheme, and the corresponding Hamiltonian matrix is explicitly constructed and diagonalized to obtain the low-lying eigenvalues. The Adam optimizer is then used to minimize the discrepancy between the predicted spectrum and the experimental data, with training typically converging within $10^4$ iterations.

To verify the validity of the reconstruction, we apply the obtained dilaton profile to the calculation of the heavy-quark spectrum and compare the results with experimental data. The results show that MLPs are highly effective at solving inverse problems and provide accurate numerical solutions. This demonstrates that neural networks can serve as a reliable tool for data-driven holographic modeling.

We acknowledge that reconstructing the dilaton field is an inverse problem, and, in principle, multiple dilaton profiles may reproduce the same spectrum and decay constants within the error margins. To examine the robustness of our approach, we tested different activation functions (SiLU, ReLU, and Tanh) combined with error-band analysis, see Figure \ref{fig:4}. This small-scale experiment exposes a subtle but important caveat: \emph{when using neural networks as interpolators for sparse hadron spectroscopy data, the choice of activation function is not merely a technical detail but encodes a strong inductive bias}. Bounded, saturating activations such as Tanh introduce a natural safeguard against unphysical extrapolation, an insight that aligns with recent theoretical work on the extrapolation behavior of deep networks. For larger datasets, the difference may wash out, but in the common case of a handful of excited states, selecting a function with built–in regularity is a simple and effective alternative to hand-tuned explicit regularisation. These findings encourage the consideration of activation design as an integral part of model regularisation in physics-informed neural networks.

The physical significance of this work is that the dilaton profile reconstructed by the MLP provides a unified, data-driven holographic framework that can simultaneously reproduce the mass spectrum and decay constants of heavy quarkonia. This achievement is not merely a numerical fit, but reveals the underlying dynamical structure of heavy-quark systems: on the one hand, the confinement mechanism of the strong interaction naturally emerges within the holographic description; on the other hand, the suppression of the wave function at the origin governs the systematic decrease of decay constants with radial excitation. In other words, the MLP profile unifies these seemingly independent physical phenomena into a single holographic potential, highlighting the ability of holographic QCD to capture the essential dynamics of heavy quarkonia.

The findings not only provide an effective paradigm for applying neural networks to 
holographic QCD but also point to new directions for subsequent research. These directions include a more systematic exploration of the relation between the reconstructed background fields and various hadronic observables, as well as the integration of data-driven methods with analytical holographic models, aiming to achieve a more comprehensive understanding of the nonperturbative dynamics of QCD. 

It should be emphasized that the present results mainly focus on the zero-temperature spectrum. On the one hand, the framework can be naturally extended to finite-temperature backgrounds to investigate spectral functions and melting phenomena. On the other hand, this work may inspire future efforts to incorporate additional experimental inputs, such as meson spectra in other channels, decay widths, or the equation of state, into the holographic reconstruction. Ultimately, such developments may enable the construction of a unified data-driven holographic framework capable of describing broader aspects of strong-interaction physics.

While the model achieves excellent agreement with existing data, its predictive power for higher radial excitations ($n>6$) remains untested due to limited experimental inputs. Furthermore, the current framework does not incorporate open-flavor thresholds or coupled-channel effects, which may influence higher-lying states. Future extensions could integrate lattice QCD data, finite-temperature backgrounds, or external magnetic fields to explore thermal spectral functions and magnetized quarkonia. Incorporating additional observables such as decay widths and form factors could further constrain the dilaton profile and enhance the universality of data-driven dilaton-based models.

\section*{Data Availability Statement}
The data that support the findings of this article are openly available [94].

\section*{Acknowledgement}
    This work is supported by National Natural Science Foundation of China (NSFC) Grants No. 12405154 (X. Chen), 12350410371 (M.A. Martin Contreras), and the European Union -- Next Generation EU through the research grant number P2022Z4P4B ``SOPHYA - Sustainable Optimised PHYsics Algorithms: fundamental physics to build an advanced society'' under the program PRIN 2022 PNRR of the Italian Ministero dell'Universit\`a e Ricerca (MUR).
\section*{References}
    
    \bibliography{references}  

@article{Polchinski:2001tt,
  author         = "Polchinski, Joseph and Strassler, Matthew J.",
  title          = "{Hard scattering and gauge/string duality}",
  journal        = "Phys. Rev. Lett.",
  volume         = "88",
  pages          = "031601",
  year           = "2002",
  eprint         = "hep-th/0109174",
  archivePrefix  = "arXiv",
  doi            = "10.1103/PhysRevLett.88.031601"
}

@article{BoschiFilho:2002vd,
  author         = "Boschi-Filho, Henrique and Braga, Nelson R. F.",
  title          = "{Gauge/string duality and scalar glueball mass ratios}",
  journal        = "JHEP",
  volume         = "05",
  pages          = "009",
  year           = "2003",
  eprint         = "hep-th/0212207",
  archivePrefix  = "arXiv",
  doi            = "10.1088/1126-6708/2003/05/009"
}

@article{Erlich:2005qh,
  author         = "Erlich, Joshua and Katz, Emanuel and Son, Dam T. and Stephanov, Mikhail A.",
  title          = "{QCD and a Holographic Model of Hadrons}",
  journal        = "Phys. Rev. Lett.",
  volume         = "95",
  pages          = "261602",
  year           = "2005",
  eprint         = "hep-ph/0501128",
  archivePrefix  = "arXiv",
  doi            = "10.1103/PhysRevLett.95.261602"
}

@article{Brodsky:2007hb,
  author         = "Brodsky, Stanley J. and de Teramond, Guy F.",
  title          = "{Light-front dynamics and AdS/QCD correspondence: The pion form factor in the space- and time-like regions}",
  journal        = "Phys. Rev. D",
  volume         = "77",
  pages          = "056007",
  year           = "2008",
  eprint         = "0707.3859",
  archivePrefix  = "arXiv",
  primaryClass   = "hep-ph",
  doi            = "10.1103/PhysRevD.77.056007"
}

@article{Maldacena:1997re,
  author         = "Maldacena, Juan Martin",
  title          = "{The Large N Limit of Superconformal Field Theories and Supergravity}",
  journal        = "Int. J. Theor. Phys.",
  volume         = "38",
  pages          = "1113--1133",
  year           = "1999",
  eprint         = "hep-th/9711200",
  archivePrefix  = "arXiv",
  doi            = "10.1023/A:1026654312961"
}

@article{Aharony:1999ti,
  author         = "Aharony, Ofer and Gubser, Steven S. and Maldacena, Juan Martin and Ooguri, Hirosi and Oz, Yaron",
  title          = "{Large N Field Theories, String Theory and Gravity}",
  journal        = "Phys. Rept.",
  volume         = "323",
  pages          = "183--386",
  year           = "2000",
  eprint         = "hep-th/9905111",
  archivePrefix  = "arXiv",
  doi            = "10.1016/S0370-1573(99)00083-6"
}

@article{Gubser:1998bc,
  author        = {Gubser, S. S. and Klebanov, I. R. and Polyakov, A. M.},
  title         = {Gauge theory correlators from non-critical string theory},
  journal       = {Phys. Lett. B},
  volume        = {428},
  pages         = {105--114},
  year          = {1998},
  eprint        = {hep-th/9802109},
  archivePrefix = {arXiv},
  doi           = {10.1016/S0370-2693(98)00377-3}
}

@article{Karch:2002sh,
  author         = "Karch, Andreas and Katz, Emanuel",
  title          = "{Adding flavor to AdS/CFT}",
  journal        = "JHEP",
  volume         = "06",
  pages          = "043",
  year           = "2002",
  eprint         = "hep-th/0205236",
  archivePrefix  = "arXiv",
  doi            = "10.1088/1126-6708/2002/06/043"
}

@article{Sakai:2004cn,
  author         = "Sakai, Tadakatsu and Sugimoto, Shigeki",
  title          = "{Low energy hadron physics in holographic QCD}",
  journal        = "Prog. Theor. Phys.",
  volume         = "113",
  pages          = "843--882",
  year           = "2005",
  eprint         = "hep-th/0412141",
  archivePrefix  = "arXiv",
  doi            = "10.1143/PTP.113.843"
}

@article{Sakai:2005yt,
  author         = "Sakai, Tadakatsu and Sugimoto, Shigeki",
  title          = "{More on a holographic dual of QCD}",
  journal        = "Prog. Theor. Phys.",
  volume         = "114",
  pages          = "1083--1118",
  year           = "2005",
  eprint         = "hep-th/0507073",
  archivePrefix  = "arXiv",
  doi            = "10.1143/PTP.114.1083"
}

@article{BoschiFilho:2002ta,
  author         = "Boschi-Filho, Henrique and Braga, Nelson R. F.",
  title          = "{QCD/String holographic mapping and glueball mass spectrum}",
  journal        = "Eur. Phys. J. C",
  volume         = "32",
  pages          = "529--533",
  year           = "2004",
  eprint         = "hep-th/0209080",
  archivePrefix  = "arXiv",
  doi            = "10.1140/epjc/s2003-01426-9"
}

@article{Grigoryan:2007my,
  author         = "Grigoryan, Hovhannes R. and Radyushkin, Anatoly V.",
  title          = "{Structure of vector mesons in a holographic model with linear confinement}",
  journal        = "Phys. Rev. D",
  volume         = "76",
  pages          = "095007",
  year           = "2007",
  eprint         = "0706.1543",
  archivePrefix  = "arXiv",
  primaryClass   = "hep-ph",
  doi            = "10.1103/PhysRevD.76.095007"
}

@article{Erdmenger:2007cm,
  author         = "Erdmenger, Johanna and Evans, Nick and Kirsch, Ingo and Threlfall, Ed",
  title          = "{Mesons in Gauge/Gravity Duals - A Review}",
  journal        = "Eur. Phys. J. A",
  volume         = "35",
  pages          = "81--133",
  year           = "2008",
  eprint         = "0711.4467",
  archivePrefix  = "arXiv",
  primaryClass   = "hep-th",
  doi            = "10.1140/epja/i2007-10540-1"
}

@article{Colangelo:2008us,
  author         = "Colangelo, Pietro and De Fazio, Fulvia and Giannuzzi, Floriana and Jugeau, Frederic and Nicotri, Stefano",
  title          = "{Light scalar mesons in the soft-wall model of AdS/QCD}",
  journal        = "Phys. Rev. D",
  volume         = "78",
  pages          = "055009",
  year           = "2008",
  eprint         = "0807.1054",
  archivePrefix  = "arXiv",
  primaryClass   = "hep-ph",
  doi            = "10.1103/PhysRevD.78.055009"
}

@article{BallonBayona:2009ar,
  author         = "Ballon Bayona, Alfonso and Boschi-Filho, Henrique and Braga, Nelson R. F. and Torres, Marcus A. C.",
  title          = "{Deep inelastic scattering for vector mesons in holographic D4-D8 model}",
  journal        = "JHEP",
  volume         = "01",
  pages          = "052",
  year           = "2010",
  eprint         = "0911.0023",
  archivePrefix  = "arXiv",
  primaryClass   = "hep-th",
  doi            = "10.1007/JHEP01(2010)052"
}

@article{Cotrone:2010fy,
  author         = "Cotrone, Aldo L. and Dymarsky, Anatoly and Kuperstein, Stanislav",
  title          = "{On Vector Meson Masses in a Holographic SQCD}",
  journal        = "JHEP",
  volume         = "03",
  pages          = "005",
  year           = "2011",
  eprint         = "1010.1017",
  archivePrefix  = "arXiv",
  primaryClass   = "hep-th",
  doi            = "10.1007/JHEP03(2011)005"
}

@article{Kim:2007zzm,
  author         = "Kim, Youngman and Lee, Jong-Phil and Lee, Su Houng",
  title          = "{Heavy quarkonium in a holographic QCD model}",
  journal        = "Phys. Rev. D",
  volume         = "75",
  pages          = "114008",
  year           = "2007",
  eprint         = "hep-ph/0703172",
  archivePrefix  = "arXiv",
  primaryClass   = "hep-ph",
  doi            = "10.1103/PhysRevD.75.114008"
}

@article{Grigoryan:2010pj,
  author         = "Grigoryan, Hovhannes R. and Hohler, Paul M. and Stephanov, Mikhail A.",
  title          = "{Towards the Gravity Dual of Quarkonium in the Strongly Coupled QCD Plasma}",
  journal        = "Phys. Rev. D",
  volume         = "82",
  pages          = "026005",
  year           = "2010",
  eprint         = "1003.1138",
  archivePrefix  = "arXiv",
  primaryClass   = "hep-ph",
  doi            = "10.1103/PhysRevD.82.026005"
}

@article{Li:2015zda,
  author         = "Li, Yang and Maris, Pieter and Zhao, Xingbo and Vary, James P.",
  title          = "{Heavy Quarkonium in a Holographic Basis Light-Front Quantization Approach}",
  journal        = "Phys. Lett. B",
  volume         = "758",
  pages          = "118--124",
  year           = "2016",
  eprint         = "1509.07212",
  archivePrefix  = "arXiv",
  primaryClass   = "hep-ph",
  doi            = "10.1016/j.physletb.2016.04.065"
}

@article{MartinContreras:2021bis,
    author = "Martin Contreras, Miguel Angel and Diles, Saulo and Vega, Alfredo",
    title = "{Heavy quarkonia spectroscopy at zero and finite temperature in bottom-up AdS/QCD}",
    eprint = "2101.06212",
    archivePrefix = "arXiv",
    primaryClass = "hep-ph",
    doi = "10.1103/PhysRevD.103.086008",
    journal = "Phys. Rev. D",
    volume = "103",
    number = "8",
    pages = "086008",
    year = "2021"
}

@article{Navas:2024,
  author         = "Navas, S. and others",
  collaboration  = "Particle Data Group",
  title          = "{Review of Particle Physics}",
  journal        = "Phys. Rev. D",
  volume         = "110",
  pages          = "030001",
  year           = "2024",
  doi            = "10.1103/PhysRevD.110.030001"
}

@article{Evans:2006ea,
  author        = {Evans, Nick and Tedder, Andrew},
  title         = {Perfecting the AdS/QCD correspondence},
  journal       = {Phys. Lett. B},
  volume        = {642},
  pages         = {546--550},
  year          = {2006},
  doi           = {10.1016/j.physletb.2006.09.060},
  eprint        = {hep-ph/0609112},
  archivePrefix = {arXiv},
  primaryClass  = {hep-ph}
}

@article{Afonin:2011ff,
  author        = {Afonin, S. S.},
  title         = {Low-energy holographic models for QCD},
  journal       = {Phys. Rev. C},
  volume        = {83},
  pages         = {048202},
  year          = {2011},
  doi           = {10.1103/PhysRevC.83.048202},
  eprint        = {1102.0156},
  archivePrefix = {arXiv},
  primaryClass  = {hep-ph}
}

@article{Liu:2026triquark,
  author       = {Xin-Ru Liu and Qi Liu and He Song and Chen},
  title        = {Regge trajectories for the doubly heavy triquarks $((Qq)\bar{Q}')$},
  journal      = {arXiv preprint},
  eprint       = {2602.16988},
  archivePrefix= {arXiv},
  primaryClass = {hep-ph},
  year         = {2026},
  month        = {Feb},
}

@article{Afonin:2012jn,
  author        = {Afonin, S. S.},
  title         = {Soft wall model with inverse exponential profile as a model for the axial and pseudoscalar mesons},
  journal       = {Int. J. Mod. Phys. A},
  volume        = {27},
  pages         = {1250171},
  year          = {2012},
  doi           = {10.1142/S0217751X12501710},
  eprint        = {1207.2644},
  archivePrefix = {arXiv},
  primaryClass  = {hep-ph}
}

@article{Braga:2015jca,
  author        = {Braga, Nelson R. F. and Martin Contreras, Miguel Angel and Diles, Saulo},
  title         = {Decay constants in soft wall AdS/QCD revisited},
  journal       = {Phys. Lett. B},
  volume        = {763},
  pages         = {203--207},
  year          = {2016},
  doi           = {10.1016/j.physletb.2016.10.046},
  eprint        = {1507.04708},
  archivePrefix = {arXiv},
  primaryClass  = {hep-ph}
}

@article{Braga:2018hjt,
  author        = {Braga, Nelson R. F. and Ferreira, Luiz F.},
  title         = {Holographic description of heavy vector mesons in a finite density plasma},
  journal       = {Phys. Lett. B},
  volume        = {783},
  pages         = {186--190},
  year          = {2018},
  doi           = {10.1016/j.physletb.2018.06.057},
  eprint        = {1802.02084},
  archivePrefix = {arXiv},
  primaryClass  = {hep-ph}
}

@article{Ma:2023ml,
  author        = {Ma, Yu-Gang and Pang, Long-Gang and Wang, Rui and Zhou, Kai},
  title         = {Phase Transition Study Meets Machine Learning},
  journal       = {Chin. Phys. Lett.},
  volume        = {40},
  pages         = {122101},
  year          = {2023},
  doi           = {10.1088/0256-307X/40/12/122101},
  eprint        = {2311.07274},
  archivePrefix = {arXiv},
  primaryClass  = {nucl-th}
}

@inproceedings{Zhou:2023fair,
  author        = {Zhou, Kai and Pang, Long-Gang and Shi, Shuzhe and Stoecker, Horst},
  title         = {Deep Learning for inverse problems in nuclear physics},
  booktitle     = {Proceedings of Science, FAIRness2022},
  pages         = {064},
  year          = {2023},
  doi           = {10.22323/1.419.0064}
}

@article{Pang:2024ijmpe,
  author        = {Pang, Long-Gang},
  title         = {Artificial Intelligence for High-Energy Nuclear Physics},
  journal       = {Int. J. Mod. Phys. E},
  volume        = {33},
  pages         = {2430009},
  year          = {2024},
  doi           = {10.1142/S0218301324300090}
}

@article{Li:2023pnn,
  author        = {Li, Zhi-Hui and Li, Chao-Qiang and Pang, Long-Gang},
  title         = {Physics-Informed Neural Networks for Gravitational Wave Surrogate Modeling},
  year          = {2023},
  eprint        = {2309.07397},
  archivePrefix = {arXiv},
  primaryClass  = {gr-qc}
}

@article{Pu:2023dl,
  author        = {Pu, Ke-Fan and Li, Hao and Lu, Hong-Liang and Pang, Long-Gang},
  title         = {Applications of deep learning in relativistic heavy-ion collisions},
  journal       = {Chin. Phys. C},
  volume        = {47},
  pages         = {054104},
  year          = {2023},
  doi           = {10.1088/1674-1137/acbf23},
  eprint        = {2303.03934},
  archivePrefix = {arXiv},
  primaryClass  = {nucl-th}
}

@article{Pang:2021npa,
  author        = {Pang, Long-Gang},
  title         = {Machine learning for high energy heavy ion collisions},
  journal       = {Nucl. Phys. A},
  volume        = {1005},
  pages         = {121972},
  year          = {2021},
  doi           = {10.1016/j.nuclphysa.2020.121972}
}

@article{Steinheimer:2019iso,
  author        = {Steinheimer, J. and Pang, L.-G. and Zhou, K. and Koch, V. and Randrup, J. and Stoecker, H.},
  title         = {A machine learning study to identify spinodal clumping in high energy nuclear collisions},
  journal       = {JHEP},
  volume        = {12},
  pages         = {122},
  year          = {2019},
  doi           = {10.1007/JHEP12(2019)122},
  eprint        = {1906.06562},
  archivePrefix = {arXiv},
  primaryClass  = {nucl-th}
}

@article{Wang:2023lightning,
  author        = {Wang, Lingxiao and Hare, Brian M. and Zhou, Kai and Stoecker, Horst and Scholten, Olaf},
  title         = {Algorithms illuminate lightning structures},
  journal       = {Chaos, Solitons \& Fractals},
  volume        = {170},
  pages         = {113346},
  year          = {2023},
  doi           = {10.1016/j.chaos.2023.113346}
}

@article{Huang:2025ml,
  author        = {Huang, Y. and Chen, J. and Jia, J. and Liu, L.-M. and Ma, Y.-G. and Zhang, C.},
  title         = {Machine learning assisted nuclear physics studies},
  year          = {2025},
  eprint        = {2501.01352},
  archivePrefix = {arXiv},
  primaryClass  = {nucl-th}
}

@article{He:2023ml,
  author        = {He, Wan-Bing and Ma, Yu-Gang and Pang, Long-Gang and Song, Huichao and Zhou, Kai},
  title         = {Exploring QCD phase structure with machine learning},
  journal       = {Nucl. Sci. Tech.},
  volume        = {34},
  pages         = {88},
  year          = {2023},
  doi           = {10.1007/s41365-023-01133-5},
  eprint        = {2303.06752},
  archivePrefix = {arXiv},
  primaryClass  = {hep-ph}
}

@article{He:2021ml,
  author        = {He, Junjie and He, Wan-Bing and Ma, Yu-Gang and Zhang, Song},
  title         = {Machine-learning-based identification for initial clustering structure in relativistic heavy-ion collisions},
  journal       = {Phys. Rev. C},
  volume        = {104},
  pages         = {044902},
  year          = {2021},
  doi           = {10.1103/PhysRevC.104.044902},
  eprint        = {2109.06277},
  archivePrefix = {arXiv},
  primaryClass  = {hep-ph}
}

@article{Hornik:1989,
  author  = {Hornik, Kurt and Stinchcombe, Maxwell and White, Harold},
  title   = {Multilayer feedforward networks are universal approximators},
  journal = {Neural Networks},
  volume  = {2},
  number  = {5},
  pages   = {359--366},
  year    = {1989},
  doi     = {10.1016/0893-6080(89)90020-8}
}

@article{Raissi:2019pinn,
  author  = {Raissi, Maziar and Perdikaris, Paris and Karniadakis, George E.},
  title   = {Physics-informed neural networks: A deep learning framework for solving forward and inverse problems involving nonlinear partial differential equations},
  journal = {Journal of Computational Physics},
  volume  = {378},
  pages   = {686--707},
  year    = {2019},
  doi     = {10.1016/j.jcp.2018.10.045}
}

@article{Soma:2023eos,
  author        = {Soma, Shriya and Wang, Lingxiao and Shi, Shuzhe and Stoecker, Horst and Zhou, Kai},
  title         = {Reconstructing the neutron star equation of state from observational data via automatic differentiation},
  journal       = {Phys. Rev. D},
  volume        = {107},
  pages         = {083028},
  year          = {2023},
  doi           = {10.1103/PhysRevD.107.083028},
  eprint        = {2209.08883},
  archivePrefix = {arXiv},
  primaryClass  = {astro-ph.HE}
}

@article{Tong:2026MesonNN,
  author       = {Xin Tong and Wei Feng and Weiwei Xu and Chao-Hsi Chang and Guo-Li Wang and Qiang Li},
  title        = {Meson Properties and Symmetry Emergence Based on the Deep Neural Network},
  journal      = {Chinese Physics Letters},
  year         = {2026},
  volume       = {43},
  number       = {2},
  pages        = {020201},
  doi          = {10.1088/0256-307X/43/2/020201}
}

@article{Mei:2025,
  author        = {Jie Mei and Lingxiao Wang and Mei Huang},
  title         = {Neural network extraction of chromo-electric and chromo-magnetic gluon masses},
  eprint        = {2507.22012},
  archivePrefix = {arXiv},
  primaryClass  = {hep-ph},
  year          = {2025}
}

@article{Karniadakis:2021piml,
  author  = {Karniadakis, George Em and Kevrekidis, Ioannis G. and Lu, Lu and Perdikaris, Paris and Wang, Sifan and Yang, Liu},
  title   = {Physics-informed machine learning},
  journal = {Nature Reviews Physics},
  volume  = {3},
  pages   = {422--440},
  year    = {2021},
  doi     = {10.1038/s42254-021-00314-5}
}

@article{Shi:2022hq,
  author        = {Shi, Shuzhe and Zhou, Kai and Zhao, Jiaxing and Mukherjee, Swagato and Zhuang, Pengfei},
  title         = {Heavy quark potential in the quark-gluon plasma: Deep neural network meets lattice quantum chromodynamics},
  journal       = {Phys. Rev. D},
  volume        = {105},
  pages         = {014017},
  year          = {2022},
  doi           = {10.1103/PhysRevD.105.014017},
  eprint        = {2105.07862},
  archivePrefix = {arXiv},
  primaryClass  = {hep-ph}
}

@article{He:2023mlnuclear,
  author        = {He, Wanbing and Li, Qingfeng and Ma, Yugang and Niu, Zhongming and Pei, Junchen and Zhang, Yingxun},
  title         = {Machine learning in nuclear physics at low and intermediate energies},
  journal       = {Sci. China Phys. Mech. Astron.},
  volume        = {66},
  pages         = {282001},
  year          = {2023},
  doi           = {10.1007/s11433-023-2116-0},
  eprint        = {2301.06396},
  archivePrefix = {arXiv},
  primaryClass  = {nucl-th}
}

@inproceedings{Zhou:2018eos,
  author    = {Zhou, Kai and Pang, Long-Gang and Su, Nan and Petersen, Hannah and Stoecker, Horst and Wang, Xin-Nian},
  title     = {An equation-of-state-meter of QCD transition from deep learning},
  booktitle = {EPJ Web Conf.},
  volume    = {171},
  pages     = {16005},
  year      = {2018},
  doi       = {10.1051/epjconf/201817116005}
}

@article{Pang:2019eos,
  author    = {Pang, Long-Gang and Zhou, Kai and Su, Nan and Petersen, Hannah and Stoecker, Horst and Wang, Xin-Nian},
  title     = {An equation-of-state-meter of QCD transition from deep learning},
  journal   = {Nucl. Phys. A},
  volume    = {982},
  pages     = {867--870},
  year      = {2019},
  doi       = {10.1016/j.nuclphysa.2018.09.095}
}

@article{Wang:2022piml,
  author        = {Wang, Lei and Jiang, Yi and He, Liang and Zhou, Kai},
  title         = {Physics-informed machine learning for discovering order parameters in phase transitions},
  journal       = {Chin. Phys. Lett.},
  volume        = {39},
  pages         = {120502},
  year          = {2022},
  doi           = {10.1088/0256-307X/39/12/120502},
  eprint        = {2005.04857},
  archivePrefix = {arXiv},
  primaryClass  = {cond-mat.dis-nn}
}

@article{Zhao:2022bnn,
  author        = {Zhao, Ya-Song and Wang, Lingxiao and Zhou, Kai and Huang, Xu-Guang},
  title         = {Bayesian neural network study of the neutron star equation of state from gravitational wave observations},
  journal       = {Phys. Rev. C},
  volume        = {106},
  pages         = {L051901},
  year          = {2022},
  doi           = {10.1103/PhysRevC.106.L051901},
  eprint        = {2105.13761},
  archivePrefix = {arXiv},
  primaryClass  = {hep-ph}
}

@article{Du:2020dljet,
  author        = {Du, Yi-Lun and Pablos, Daniel and Tywoniuk, Konrad},
  title         = {Deep learning jet modifications in heavy-ion collisions},
  journal       = {JHEP},
  volume        = {2021},
  pages         = {206},
  year          = {2020},
  doi           = {10.1007/JHEP21(2020)206},
  eprint        = {2012.07797},
  archivePrefix = {arXiv},
  primaryClass  = {hep-ph}
}

@article{Du:2022jettomo,
  author        = {Du, Yi-Lun and Pablos, Daniel and Tywoniuk, Konrad},
  title         = {Jet tomography in heavy-ion collisions with deep learning},
  journal       = {Phys. Rev. Lett.},
  volume        = {128},
  pages         = {012301},
  year          = {2022},
  doi           = {10.1103/PhysRevLett.128.012301},
  eprint        = {2106.11271},
  archivePrefix = {arXiv},
  primaryClass  = {hep-ph}
}

@article{Jiang:2021eos,
  author        = {Jiang, Li and Wang, Lingxiao and Zhou, Kai},
  title         = {Identifying the QCD equation of state in relativistic heavy-ion collisions with deep learning},
  journal       = {Phys. Rev. D},
  volume        = {103},
  pages         = {116023},
  year          = {2021},
  doi           = {10.1103/PhysRevD.103.116023},
  eprint        = {2103.04090},
  archivePrefix = {arXiv},
  primaryClass  = {nucl-th}
}

@inproceedings{Shi:2022lattice,
  author    = {Shi, Shuzhe and Zhou, Kai and Zhao, Jiaxing and Mukherjee, Swagato and Zhuang, Pengfei},
  title     = {Heavy quark potential in the quark--gluon plasma from deep learning},
  booktitle = {PoS LATTICE2021},
  pages     = {537},
  year      = {2022},
  doi       = {10.22323/1.396.0537}
}

@article{Hashimoto:2018dlholo,
  author        = {Hashimoto, Koji and Sugishita, Sotaro and Tanaka, Akinori and Tomiya, Akio},
  title         = {Deep Learning and Holographic QCD},
  journal       = {Phys. Rev. D},
  volume        = {98},
  pages         = {106014},
  year          = {2018},
  doi           = {10.1103/PhysRevD.98.106014},
  eprint        = {1809.10536},
  archivePrefix = {arXiv},
  primaryClass  = {hep-th}
}

@article{Yan:2020dlbh,
  author        = {Yan, Yi-Kai and Wu, Shao-Feng and Ge, Xiang-Hua and Tian, Yu},
  title         = {Deep learning black hole metrics from holographic QCD data},
  journal       = {Phys. Rev. D},
  volume        = {102},
  pages         = {101902},
  year          = {2020},
  doi           = {10.1103/PhysRevD.102.101902},
  eprint        = {2004.12112},
  archivePrefix = {arXiv},
  primaryClass  = {hep-th}
}

@article{Hashimoto:2022dilaton,
  author        = {Hashimoto, Koji and Ohashi, Keisuke and Sumimoto, Takayuki},
  title         = {Deriving dilaton potential in improved holographic QCD from meson spectrum},
  journal       = {Phys. Rev. D},
  volume        = {105},
  pages         = {106008},
  year          = {2022},
  doi           = {10.1103/PhysRevD.105.106008},
  eprint        = {2108.08091},
  archivePrefix = {arXiv},
  primaryClass  = {hep-th}
}

@article{Song:2021adsdbm,
  author        = {Song, Minyong and Oh, Min-Seok H. and Ahn, Y. and Kim, Keun-Young},
  title         = {AdS/CFT correspondence as a deep Boltzmann machine},
  journal       = {Chin. Phys. C},
  volume        = {45},
  pages         = {073111},
  year          = {2021},
  doi           = {10.1088/1674-1137/abf6b0},
  eprint        = {2011.13726},
  archivePrefix = {arXiv},
  primaryClass  = {physics.class-ph}
}

@article{Chang:2024hqanisotropy,
  author        = {Chang, Wen-Bin and Hou, De-fu},
  title         = {Heavy quarkonium spectral function in an anisotropic background},
  journal       = {Phys. Rev. D},
  volume        = {109},
  pages         = {086010},
  year          = {2024},
  doi           = {10.1103/PhysRevD.109.086010},
  eprint        = {2403.04966},
  archivePrefix = {arXiv},
  primaryClass  = {hep-ph}
}

@article{Ahn:2024dlconductivity,
  author        = {Ahn, Byoungjoon and Jeong, Hyun-Sik and Kim, Keun-Young and Yun, Kwan},
  title         = {Deep learning bulk spacetime from boundary optical conductivity},
  journal       = {JHEP},
  volume        = {2024},
  number        = {03},
  pages         = {141},
  year          = {2024},
  doi           = {10.1007/JHEP03(2024)141},
  eprint        = {2401.00939},
  archivePrefix = {arXiv},
  primaryClass  = {hep-th}
}

@article{Gu:2024dlee,
  author        = {Gu, Zi-Fan and Yan, Yu-Kai and Wu, Shao-Feng},
  title         = {Deep learning holographic entanglement entropy},
  year          = {2024},
  eprint        = {2401.09946},
  archivePrefix = {arXiv},
  primaryClass  = {hep-th}
}

@article{Li:2023bhmetric,
  author        = {Li, Kai and Ling, Yi and Liu, Peng and Wu, Meng-He},
  title         = {Learning the black hole metric from holographic conductivity},
  journal       = {Phys. Rev. D},
  volume        = {107},
  pages         = {066021},
  year          = {2023},
  doi           = {10.1103/PhysRevD.107.066021},
  eprint        = {2209.05203},
  archivePrefix = {arXiv},
  primaryClass  = {hep-th}
}

@article{Cai:2024qcdml,
  author        = {Cai, Rong-Gen and He, Song and Li, Li and Zeng, Hong-An},
  title         = {QCD Phase Diagram at Finite Magnetic Field and Chemical Potential: A Holographic Approach Using Machine Learning},
  year          = {2024},
  eprint        = {2406.12772},
  archivePrefix = {arXiv},
  primaryClass  = {hep-th}
}

@article{Ahn:2024bhrecon,
  author        = {Ahn, Byoungjoon and Jeong, Hyun-Sik and Kim, Keun-Young and Yun, Kwan},
  title         = {Holographic reconstruction of black hole spacetime: machine learning and entanglement entropy},
  year          = {2024},
  eprint        = {2406.07395},
  archivePrefix = {arXiv},
  primaryClass  = {hep-th}
}

@article{Akutagawa:2020dladsqcd,
  author        = {Akutagawa, Tetsuya and Hashimoto, Koji and Sumimoto, Takayuki},
  title         = {Deep Learning and AdS/QCD},
  journal       = {Phys. Rev. D},
  volume        = {102},
  pages         = {026020},
  year          = {2020},
  doi           = {10.1103/PhysRevD.102.026020},
  eprint        = {2005.02636},
  archivePrefix = {arXiv},
  primaryClass  = {hep-th}
}

@article{Karch:2006pv,
  author         = {Karch, Andreas and Katz, Emanuel and Son, Dam T. and Stephanov, Mikhail A.},
  title          = {Linear Confinement and AdS/QCD},
  journal        = {Phys. Rev. D},
  volume         = {74},
  pages          = {015005},
  year           = {2006},
  doi            = {10.1103/PhysRevD.74.015005},
  eprint         = {hep-ph/0602229},
  archivePrefix  = {arXiv},
  primaryClass   = {hep-ph}
}

@article{Batell:2008zm,
  author         = {Batell, Brian and Gherghetta, Tony},
  title          = {Dynamical Soft-Wall AdS/QCD},
  journal        = {Phys. Rev. D},
  volume         = {78},
  pages          = {026002},
  year           = {2008},
  doi            = {10.1103/PhysRevD.78.026002},
  eprint         = {0801.4383},
  archivePrefix  = {arXiv},
  primaryClass   = {hep-ph}
}

@article{ParticleDataGroup:2024,
  author       = {Navas, S. and others},
  collaboration= {Particle Data Group},
  title        = {Review of Particle Physics},
  journal      = {Phys. Rev. D},
  volume       = {110},
  pages        = {030001},
  year         = {2024},
  doi          = {10.1103/PhysRevD.110.030001}
}

@article{Chen:2018bbr,
    author = "Chen, Jiao-Kai",
    title = "{Concavity of the meson Regge trajectories}",
    eprint = "1807.11003",
    archivePrefix = "arXiv",
    primaryClass = "hep-ph",
    doi = "10.1016/j.physletb.2018.10.022",
    journal = "Phys. Lett. B",
    volume = "786",
    pages = "477--484",
    year = "2018"
}

@article{Chen:2021kfw,
    author = "Chen, Jiao-Kai",
    title = "{Structure of the meson Regge trajectories}",
    eprint = "2102.07993",
    archivePrefix = "arXiv",
    primaryClass = "hep-ph",
    doi = "10.1140/epja/s10050-021-00502-y",
    journal = "Eur. Phys. J. A",
    volume = "57",
    number = "7",
    pages = "238",
    year = "2021"
}

@article{Chen:2023djq,
    author = "Chen, Jiao-Kai",
    title = "{Regge trajectory relation for the universal description of the heavy-heavy systems: Diquarks, mesons, baryons and tetraquarks}",
    eprint = "2302.05926",
    archivePrefix = "arXiv",
    primaryClass = "hep-ph",
    doi = "10.1016/j.nuclphysa.2024.122927",
    journal = "Nucl. Phys. A",
    volume = "1050",
    pages = "122927",
    year = "2024"
}

@article{MartinContreras:2020cyg,
    author = "Martin Contreras, Miguel Angel and Vega, Alfredo",
    title = "{Nonlinear Regge trajectories with AdS/QCD}",
    eprint = "2004.10286",
    archivePrefix = "arXiv",
    primaryClass = "hep-ph",
    doi = "10.1103/PhysRevD.102.046007",
    journal = "Phys. Rev. D",
    volume = "102",
    number = "4",
    pages = "046007",
    year = "2020"
}

@article{MartinContreras:2025wnh,
    author = "Martin Contreras, Miguel Angel and Vega, Alfredo",
    title = "{Engineering Confining Dilatons: A WKB Inverse Problem in Holographic QCD}",
    eprint = "2509.04956",
    archivePrefix = "arXiv",
    primaryClass = "hep-ph",
    month = "9",
    journal = {arXiv preprint},
    year = "2025"
}

@article{Braga:2018fyc,
    author = "Braga, Nelson R. F. and Ferreira, Luiz F. and Da Rocha, Rold{\~a}o",
    title = "{Thermal dissociation of heavy mesons and configurational entropy}",
    eprint = "1808.10499",
    archivePrefix = "arXiv",
    primaryClass = "hep-ph",
    doi = "10.1016/j.physletb.2018.10.036",
    journal = "Phys. Lett. B",
    volume = "787",
    pages = "16--22",
    year = "2018"
}

@article{MartinContreras:2023oqs,
    author = "Martin Contreras, Miguel Angel and Vega, Alfredo",
    title = "{Holographic stability for non-qq{\textasciimacron} candidates}",
    eprint = "2309.02905",
    archivePrefix = "arXiv",
    primaryClass = "hep-ph",
    doi = "10.1103/PhysRevD.108.126024",
    journal = "Phys. Rev. D",
    volume = "108",
    number = "12",
    pages = "126024",
    year = "2023"
}

@article{MartinContreras:2019kah,
    author = "Martin Contreras, Miguel Angel and Vega, Alfredo",
    title = "{Different approach to decay constants in AdS/QCD models}",
    eprint = "1910.10922",
    archivePrefix = "arXiv",
    primaryClass = "hep-th",
    doi = "10.1103/PhysRevD.101.046009",
    journal = "Phys. Rev. D",
    volume = "101",
    number = "4",
    pages = "046009",
    year = "2020"
}

@article{MartinContreras:2023eft,
    author = "Martin Contreras, Miguel Angel and Vega, Alfredo and Diles, Saulo",
    title = "{Isospectrality and configurational entropy as testing tools for bottom-up AdS/QCD}",
    eprint = "2308.16007",
    archivePrefix = "arXiv",
    primaryClass = "hep-ph",
    doi = "10.1016/j.physletb.2024.138723",
    journal = "Phys. Lett. B",
    volume = "854",
    pages = "138723",
    year = "2024"
}

@article{Boschi-Filho:2002xih,
    author = "Boschi-Filho, Henrique and Braga, Nelson R. F.",
    title = "{Gauge / string duality and scalar glueball mass ratios}",
    eprint = "hep-th/0212207",
    archivePrefix = "arXiv",
    doi = "10.1088/1126-6708/2003/05/009",
    journal = "JHEP",
    volume = "05",
    pages = "009",
    year = "2003"
}

@article{Lucha:1996ax,
    author = "Lucha, Wolfgang and Schoberl, Franz F.",
    title = "{Relativistic Coulomb problem: Analytic upper bounds on energy levels}",
    eprint = "hep-ph/9603429",
    archivePrefix = "arXiv",
    reportNumber = "HEPHY-PUB-632-96, UWTHPH-1995-23",
    doi = "10.1103/PhysRevA.54.3790",
    journal = "Phys. Rev. A",
    volume = "54",
    pages = "3790--3794",
    year = "1996"
}

@article{Polchinski:2002jw,
    author = "Polchinski, Joseph and Strassler, Matthew J.",
    title = "{Deep inelastic scattering and gauge / string duality}",
    eprint = "hep-th/0209211",
    archivePrefix = "arXiv",
    reportNumber = "NSF-ITP-02-139, UW-PT-02-23",
    doi = "10.1088/1126-6708/2003/05/012",
    journal = "JHEP",
    volume = "05",
    pages = "012",
    year = "2003"
}

@article{Afonin:2007gd,
    author = "Afonin, S. S.",
    title = "{Weinberg like sum rules revisited}",
    eprint = "0710.4921",
    archivePrefix = "arXiv",
    primaryClass = "hep-ph",
    doi = "10.1186/1754-0410-3-1",
    journal = "PMC Phys. A",
    volume = "3",
    pages = "1",
    year = "2009"
}

@article{Afonin:2004yb,
    author = "Afonin, S. S. and Andrianov, A. A. and Andrianov, V. A. and Espriu, D.",
    title = "{Matching Regge theory to the OPE}",
    eprint = "hep-ph/0403268",
    archivePrefix = "arXiv",
    reportNumber = "UB-ECM-PF-04-06",
    doi = "10.1088/1126-6708/2004/04/039",
    journal = "JHEP",
    volume = "04",
    pages = "039",
    year = "2004"
}

@inproceedings{Wise:1997sg,
    author = "Wise, Mark B.",
    title = "{Heavy quark physics: Course}",
    booktitle = "{Les Houches Summer School in Theoretical Physics, Session 68: Probing the Standard Model of Particle Interactions}",
    eprint = "hep-ph/9805468",
    archivePrefix = "arXiv",
    reportNumber = "CALT-68-2172",
    pages = "1051--1089",
    month = "7",
    year = "1997"
}

@article{Lucha:1991vn,
    author = "Lucha, W. and Schoberl, F. F. and Gromes, D.",
    title = "{Bound states of quarks}",
    doi = "10.1016/0370-1573(91)90001-3",
    journal = "Phys. Rept.",
    volume = "200",
    pages = "127--240",
    year = "1991"
}

@article{Krutov:2016uhy,
    author = "Krutov, A. F. and Polezhaev, R. G. and Troitsky, V. E.",
    title = "{The radius of the rho meson determined from its decay constant}",
    eprint = "1602.00907",
    archivePrefix = "arXiv",
    primaryClass = "hep-ph",
    doi = "10.1103/PhysRevD.93.036007",
    journal = "Phys. Rev. D",
    volume = "93",
    number = "3",
    pages = "036007",
    year = "2016"
}

@article{Carrillo-Serrano:2015uca,
    author = {Carrillo-Serrano, Manuel E. and Bentz, Wolfgang and Clo{\"e}t, Ian C. and Thomas, Anthony W.},
    title = "{$\rho$ meson form factors in a confining Nambu{\textendash}Jona-Lasinio model}",
    eprint = "1504.08119",
    archivePrefix = "arXiv",
    primaryClass = "nucl-th",
    reportNumber = "ADP-15-17-T919",
    doi = "10.1103/PhysRevC.92.015212",
    journal = "Phys. Rev. C",
    volume = "92",
    number = "1",
    pages = "015212",
    year = "2015"
}

@article{Aliev:2009gj,
    author = "Aliev, T. M. and Ozpineci, A. and Savci, M.",
    title = "{Magnetic and quadrupole moments of light spin-1 mesons in light cone QCD sum rules}",
    eprint = "0902.4627",
    archivePrefix = "arXiv",
    primaryClass = "hep-ph",
    reportNumber = "METU-PHYS-HEP-010-09",
    doi = "10.1016/j.physletb.2009.06.073",
    journal = "Phys. Lett. B",
    volume = "678",
    pages = "470--476",
    year = "2009"
}

@article{Vary:2018pmv,
    author = "Vary, James P. and others",
    title = "{Hadron Spectra, Decays and Scattering Properties Within Basis Light Front Quantization}",
    eprint = "1804.07865",
    archivePrefix = "arXiv",
    primaryClass = "nucl-th",
    doi = "10.1007/s00601-018-1356-0",
    journal = "Few Body Syst.",
    volume = "59",
    number = "4",
    pages = "56",
    year = "2018"
}

@article{Zeng:2025tcz,
    author = "Zeng, Hong-An and Wang, Lingxiao and Huang, Mei",
    title = "{HoloNet: Toward a Unified Einstein-Maxwell-Dilaton Framework of QCD}",
    eprint = "2512.06044",
    archivePrefix = "arXiv",
    primaryClass = "hep-lat",
    reportNumber = "RIKEN-iTHEMS-Report-25",
    month = "12",
    year = "2025"
}

@article{Wen:2024hgu,
    author = "Wen, Nanxiang and Cao, Xuanmin and Chao, Jingyi and Liu, Hui",
    title = "{Neutral pion masses within a hot and magnetized medium in a lattice-improved soft-wall AdS/QCD model}",
    eprint = "2402.06239",
    archivePrefix = "arXiv",
    primaryClass = "hep-th",
    doi = "10.1103/PhysRevD.109.086021",
    journal = "Phys. Rev. D",
    volume = "109",
    number = "8",
    pages = "086021",
    year = "2024"
}

@article{Cao:2021tcr,
    author = "Cao, Xuanmin and Qiu, Songyu and Liu, Hui and Li, Danning",
    title = "{Thermal properties of light mesons from holography}",
    eprint = "2102.10946",
    archivePrefix = "arXiv",
    primaryClass = "hep-ph",
    doi = "10.1007/JHEP08(2021)005",
    journal = "JHEP",
    volume = "08",
    pages = "005",
    year = "2021"
}

@article{Li:2013oda,
    author = "Li, Danning and Huang, Mei",
    title = "{Dynamical holographic QCD model for glueball and light meson spectra}",
    eprint = "1303.6929",
    archivePrefix = "arXiv",
    primaryClass = "hep-ph",
    doi = "10.1007/JHEP11(2013)088",
    journal = "JHEP",
    volume = "11",
    pages = "088",
    year = "2013"
}

@article{ThomasArun:2025uyi,
    author = "Thomas Arun, Mathew and Pal, Ritik",
    title = "{Learning holographic QCD with unflavoured meson spectra}",
    eprint = "2512.16450",
    archivePrefix = "arXiv",
    primaryClass = "hep-ph",
    month = "12",
    year = "2025"
}

@article{Xu2020sep,
  author       = {Keyulu Xu and
                  Jingling Li and
                  Mozhi Zhang and
                  Simon S. Du and
                  Ken{-}ichi Kawarabayashi and
                  Stefanie Jegelka},
  title        = {How Neural Networks Extrapolate: From Feedforward to Graph Neural
                  Networks},
  journal      = {CoRR},
  volume       = {abs/2009.11848},
  year         = {2020},
  url          = {https://arxiv.org/abs/2009.11848},
  eprinttype   = {arXiv},
  eprint       = {2009.11848},
  bibsource    = {dblp computer science bibliography, https://dblp.org}
}

@article{ziyin2020aug,
  author       = {Liu Ziyin and
                  Tilman Hartwig and
                  Masahito Ueda},
  title        = {Neural Networks Fail to Learn Periodic Functions and How to Fix It},
  journal      = {CoRR},
  volume       = {abs/2006.08195},
  year         = {2020},
  url          = {https://arxiv.org/abs/2006.08195},
  eprinttype   = {arXiv},
  eprint       = {2006.08195},
  bibsource    = {dblp computer science bibliography, https://dblp.org}
}

@misc{maennel2018aug,
      title={Gradient Descent Quantizes ReLU Network Features}, 
      author={Hartmut Maennel and Olivier Bousquet and Sylvain Gelly},
      year={2018},
      eprint={1803.08367},
      archivePrefix={arXiv},
      primaryClass={stat.ML},
      url={https://arxiv.org/abs/1803.08367}, 
}
    
\end{document}